\documentclass[12pt,preprint,authoryear,square]{aastex}
\font\caps=cmcsc10 scaled 1200
\def\HI {H\kern0.1em{\sc i}} 
\def\radm {rad m$^{-2}$} 
\def\dg{^{\circ}}
\def\deg{$^{\circ}$}

\begin{document}
\bibliographystyle{alpha}
\title{~~\\ ~~\\ VLBI Polarimetry of 177
Sources from the Caltech-Jodrell Bank Flat-spectrum Survey}
\shorttitle{CJF}
\shortauthors{Pollack \& Taylor \& Zavala}
\author{L. K. Pollack,\altaffilmark{1,2}  G. B. Taylor,\altaffilmark{1} R. T. Zavala\altaffilmark{1}}
\email{lpollack@uclink4.berkeley.edu, gtaylor@nrao.edu, rzavala@nrao.edu}
\altaffiltext{1}{National Radio Astronomy Observatory, P.O. Box 0, Socorro, NM 87801}
\altaffiltext{2}{University of California, Berkeley, Berkeley, CA  94704}


\slugcomment{Accepted to the Astrophysical Journal}

\begin{abstract}
We present VLBA observations and a statistical analysis of 5 GHz VLBI
polarimetry data from 177 sources in the Caltech-Jodrell Bank
flat-spectrum (CJF) survey.  The CJF survey, a complete,
flux-density-limited sample of 293 extragalactic radio sources, gives
us the unique opportunity to compare a broad range of source
properties for quasars, galaxies and BL Lacertae objects.  We focus
primarily on jet properties, specifically the correlation between the
jet axis angle and the polarization angle in the core and jet.  A
strong correlation is found for the electric vector polarization angle
in the cores of quasars to be perpendicular to the jet axis.  Contrary
to previous claims, no correlation is found between the jet
polarization angle and the jet axis in either quasars or BL Lac
objects.  With this large, homogeneous sample we are also able to
investigate cosmological issues and AGN evolution.
\end{abstract}

\keywords{galaxies: active -- galaxies: jets -- galaxies: nuclei -- radio continuum: 
galaxies; polarimetry}

\section{Introduction}

Studies of the polarization properties of jets on the milliarcsecond
scale can yield vital insights about their formation, 
collimation, and propagation \citep[e.g.][]{jlg00, mei01}. 
Based on results from 25 BL Lacs, 
\citet{gab00} claim a tendency for the electric vector polarization 
angle to lie along the jet axis for these objects, with about 30 percent
showing a perpendicular orientation.  This has been interpreted 
in terms of oblique and transverse shocks \citep{gab00} or as
the result of a helical magnetic field confining the jet
\citep{gab02}.

Although many individual sources and even large surveys have been
undertaken with VLBI in total intensity, the number of sources studied
with VLBI polarimetry is far less.  In the early years of VLBI this
was due in part to hardware restrictions (e.g., feeds with only a 
single available polarization), but also due to
software restrictions that made calibration difficult.  The VLBI group
at Brandeis pioneered efforts to develop techniques and software to
make VLBI polarimetry more tractable \citep{rob84,rob91}, and 
this culminated in a study of the mas polarization structure of 24 
sources from the Pearson-Readhead sample \citep{caw93}.  But 
it wasn't until the advent of the VLBA, with homogeneous, low-leakage 
antennas and feeds, that VLBI polarimetry became widely accessible.  
Full polarization calibration techniques were implemented in AIPS in 
the 1990s and in 1999 NRAO began providing regular monitoring\footnote{\tt
http://www.aoc.nrao.edu/$\sim$smyers/calibration/} of a suite of
calibrators that can be used for the absolute polarization angle
calibration \citep{tmy00}.

The Caltech-Jodrell Bank flat-spectrum (CJF) survey is a large
complete, flux-limited sample of sources imaged with VLBI, consisting
of 293 sources \citep{TayCJF}.  This survey has been used to
place constraints on gravitational lensing by 10$^5$ M$_\odot$ black
holes \citep{wil01}, to study the unbeamed synchrotron luminosities
of relativistic jets \citep{lis97}, and
recently to study the relation between the linear size of the VLBI
jet and the black hole mass \citep{cj02}.

As part of a proper-motion study of compact objects, all the sources
in the CJF sample were observed at least 3 times at intervals of
approximately 2 years~\citep{Britz98}.  The early epochs used
Global VLBI observations, but starting in 1995 the VLBA was used
exclusively since it could provide better ($u,v$) coverage and
consequently improved image fidelity.  The first observing sessions
with the VLBA were limited in bandwidth (8 MHz) and used only a single
polarization, but as the VLBA matured and its capabilities increased,
we began in 1998 to observe with 16 MHz bandwidth in full polarization
mode.  The last three epochs of CJF were observed in this mode and
provide full VLBI polarimetry of 177 sources.

Here we present a statistical analysis of the properties
of those 177 sources in CJF observed with full polarimetry.  
The sample is described in detail in \S2, and the calibration
procedures are described in \S3.  We present results in \S4, 
and discuss comparisons of source properties, and
evolution of those properties with redshift in \S5.  

We assume H$_0 = 75$ km s$^{-1}$ Mpc$^{-1}$ and $q_0$=0.5 throughout.

\section{Sample Selection}\label{SampSelection}

We present 5 GHz Very Long Baseline Interferometric (VLBI) polarimetry
observations of 177 sources in the Caltech-Jodrell Bank flat-spectrum
(CJF) survey, a complete flux-density-limited sample of 293
flat-spectrum radio sources.  The CJF sample requires that the flux
density at 4850 MHz be at least 350 mJy, ($S_{4850} \geq 350$ mJy),
and that the spectral index at 4850 MHz and 1400 MHz be at least
$-0.5$ ($\alpha^{4850}_{1400} \geq -0.5$).  Additionally, the sample
is limited by B1950 declinations and galactic latitudes such that
$\delta \geq 35\dg$ and $|\it{b}|$ $\geq 10\dg$.  A list of the 293
extragalactic radio sources and a more complete description of the
selection criteria for the CJF sample can be found in \citet{TayCJF}.
The CJF sample has been completely observed with VLBI as part of the
PR \citep{pr88}, CJ1 \citep{polcj1} and CJ2 \citep{taycj2} surveys.

The 177 sources presented here were observed as part of a proper
motion study~\citep{Britz98} so that polarimetric data in
multiple epochs are available for 66 of the 177 sources.  Where data
from multiple epochs was available, we preferred the 2000.958
observations over the 1998.122 observations, and used the 1999.890
observations, with somewhat poorer {\it u,v} coverage, only when other
epochs were unavailable.  All plots shown here include just one epoch
of each of the 177 sources.  In the future it may be interesting to
study variability in polarized intensity and polarization angle using
those 66 sources with multiple epochs available, however no effort has
been made toward this goal thus far.

Using optical classifications from the literature \citep{hen95,ver96},
from the NASA Extragalactic Database, and from unpublished
observations by the CJ collaboration, we find that the final sample of
177 sources consists of 30 galaxies, 106 quasars, 20 BL Lacertae
objects and 21 others, including empty fields, red and blue objects
and other optically unresolved objects.  We note here a few words of
caution regarding the above classifications in that low-luminosity BL
Lacs can be missidentified as quasars or radio galaxies \citep{mar95},
and variability in the emission lines of BL Lacs and quasars can cause
some movement between these classes \citep{ver95}.  Redshifts have
been measured for 147 of the sources presented here.

\section{Observations}\label{Obs}
\subsection{{Calibration}}\label{Calib}

The observations were carried out on 2000 December 16 and 17
(2000.958), 1999 November 21, 23 and 26 (1999.890), and 1998 February
8, 12, 13, 20 and 21 (1998.122).  These three epochs provided about
34, 72 and 120 hours of data, respectively, for a total of 226 hours
of observations.  The 2000.958 and 1998.122 observations used the 10
element Very Long Baseline Array\footnote{The National Radio Astronomy
Observatory is operated by Associated Universities, Inc., under
cooperative agreement with the National Science Foundation.}  (VLBA)
while the 1999.890 observations used only 8 of the VLBA antennas.  The
Saint Croix, Virgin Islands antenna was lost due to hurricane Lenny,
and the North Liberty, Iowa antenna was lost due to a power hardware
problem.  Because we did not use phase referencing, we failed to image
6 sources with total intensity peaks of less than 55 mJy.  Of these, four 
sources were identified with galaxies, and two sources had no
optical identification.  Another source, 0954+556, was omitted due to
a position error of 30 arcseconds.  Finally, the gravitationally
lensed object 0218+357 was also omitted.  These 8 sources that we did
not image have not been included in any of the distributions or
analysis, and are not included in the 177 source count.

We observed at 4995 MHz with a total bandwidth of 16 MHz.  Right- and
left-circular polarizations were recorded using 1 bit sampling at 64
Mbits/sec.  Two intermediate frequencies (IFs) of 8 MHz each were used
for right- and left- polarization.  Amplitude calibration for each
antenna was derived from measurements of the antenna gain and system
temperatures during each run.   Global
fringe fitting was performed using the AIPS task FRING, an
implementation of the \citet{sc83} algorithm.  The fringe
fitting was performed on each IF and polarization independently using
a solution interval of 2 minutes, and a point source model was
assumed. Next, a short segment of the cross hand data from the
strongly polarized calibrator 3C\,279 was fringe fitted in order to
determine the right-left delay difference, and the correction obtained
was applied to the rest of the data.  Once delay and rate solutions
were applied the data were averaged in frequency over 8 MHz.  All data
were then averaged together over a 30 second time interval.  Imaging,
editing and self-calibration were performed in {\caps Difmap} \citep{Shep95}
and AIPS.  

   Determination of the leakage terms was performed with the AIPS task
LPCAL using a strong calibrator with simple source structure such as
0923+392 or 0716+714.  Given the nature of the survey, strong sources
with good parallactic angle coverage were abundant in every observing
session.  The absolute electric vector polarization angle (EVPA)
calibration was performed using various well-known target sources such
as 0923+392, 1803+784, and 2200+420 (BL Lac).  These observations were
compared to contemporaneous short VLA observations and VLBI
polarimetry reported in the literature, or available from the VLA/VLBA
Polarization Calibration Page \citep{tmy00}.  We also used
component C4 of 3C\,279, which was fairly stable during this time
period \citep{tay00,zt01}.  The absolute uncertainty in the 
EVPA calibration is $\sim$4\deg, with the dominant source of error being
the variability of the EVPAs of the calibrator sources \citep{tmy00}.

\subsection{Data Analysis}\label{Analysis}

\begin{figure} 
\epsscale{0.87} 
\plotone{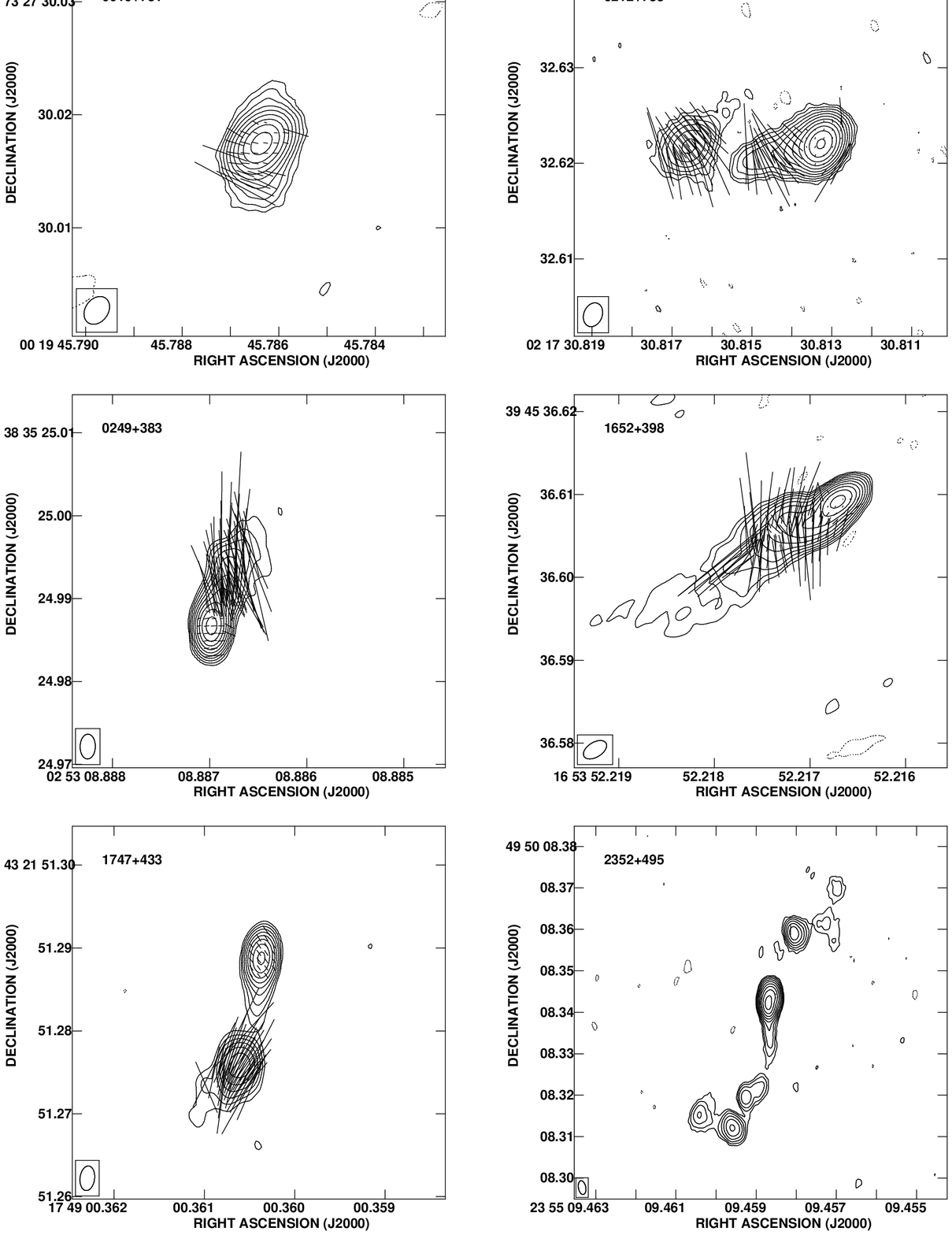}
\caption{A selection of images from the sample showing a range of
core-jet morphologies and the CSO 2352+495.  Contours are drawn
at $-4\sigma_I$, $4\sigma_I$, $8\sigma_I$, $16\sigma_I$, ..., where
$\sigma_I$ is RMS of total intensity given in Table 1.  Polarization
vectors have lengths proportional to fractional polarization.
Blanking of the polarization has been performed on pixels less than 
4.6$\sigma_p$ in polarization, 9$\sigma_I$ in total intensity, or 0.2\%
in fractional polarization.}
\label{rangemos}
\end{figure}

We automated the imaging process by writing a script that ran the AIPS
task IMAGR on the self-calibrated data for every source, creating
Stokes I, Q and U maps.  Using another script, we then ran the AIPS
task COMB on each of the sources to produce polarized intensity maps,
polarized intensity noise maps, polarization angle maps and
polarization angle noise maps.  Some sample images covering a range of
morphologies are shown in Fig.~\ref{rangemos}.  These FITS files were
brought into IDL where we used programs to objectively and uniformly
determine the properties of each source.

\subsection{{Source Property Definitions}}\label{Defs}

The RMS noise levels were calculated for the total intensity and
linear polarizations by taking the standard deviations of four corner
sections of the image and averaging the two middle values.  We
calculated the integrated flux in Stokes I, Q and U by summing the
flux in a particular region that completely enclosed each source.  The
algorithm used to systematically define the region boundaries creates
a rectangular box around those pixels with values at least 15 times
the RMS noise level.  Due to image artifacts this high cut-off proved
necessary to correctly enclose many sources.  A border of width three
times the FWHM of the convolving beam was added to the initial
rectangular area to define the final region to be used in the summing.
As a check we compared the integrated flux calculated in IDL to the
sum of the clean components found in the AIPS task IMAGR.  The total
integrated flux measurements of all 177 sources were found to agree
with that calculated by the AIPS clean algorithm to within 8\%. 

The core of each source was assumed to be located at the brightest
peak in the total intensity map when no other data were available for
study.  However, in the case of compact symmetric objects (CSOs) or
sources with complex jet structures we took advantage of 15 GHz
images \citep{trp96} (and unpublished 15 GHz observations) 
and published motion studies \citep{Tay97} to more
accurately determine core positions.  Jet component positions were
found by subtracting from the initial total intensity image a Gaussian
model centered at the peak flux position and scaled so that the
maximum value of the Gaussian equaled the peak value in the map.  As
the appropriate Gaussian is subtracted from the previous image, the
locations of each new peak define the locations of consecutive jet
components.  We add the restriction that the total intensity at the
jet component position found must be greater than 9 times the RMS
noise level, where the noise is calculated as discussed above.  At
most 8 jet component positions were required to describe a source.
The position angle of the elliptical Gaussian model used in the
subtraction was matched to the position angle of the convolving beam,
and both axes of the Gaussian were given a width of 1.5$\sigma$, where
$\sigma=\frac{FWHM}{2\sqrt{2\ln2}}$ and describes the width of the
convolving beam.  Due to slight differences in the editing process
that changed the {\it u,v} coverage for different Stokes parameters,
the beam sizes and position angles fit by the AIPS task IMAGR differed
slightly in I, Q and U.  For consistency we forced all beams to have
the dimensions and orientation of the Stokes U beam by setting the
beam parameters BMAJ, BMIN and BPA in AIPS for the Stokes Q and I
images.

The jet properties and source morphologies based on total intensities
were easily characterized after each jet component position was
defined.  We found 34 sources with no jet components which we call
{\it naked cores} and 134 sources with at least one jet component
which we call {\it core-jet} sources.  In addition, we found 9 sources
which we classified as Compact Symmetric Objects (CSOs) or CSO
candidates based on their symmetric structures.  CSOs are a recently
identified class of radio sources smaller than 1 kpc in size with
emission on both sides of the central engine, and are thought to be
very young objects ($\sim$1000 y, \citet{rea96,ows98}). 
Eight of these sources were previously identified as
CSOs or CSO candidates \citep{peck00}, and the remaining source
(0402+379) is currently under study.

We recorded the jet axis angle to be the angle defined by the core and
the closest jet component, where North is $0\dg$.  The only exception
to this is the CSO candidate 0402+379 which has a component close to
the core but seemingly unrelated to the jet.  The jet axis angle for
this source was measured by eye.  The jet length is defined in most
cases as the distance from the core to the farthest jet component from
the core, irrespective of jet bend.  It is quoted in milliarcseconds
and is uncorrected for angular distances.  In \S 5 we model our
angular size calculation after the description given in \citet{kel93},
creating a contour line at 2\% of the peak in the total intensity map.
The maximum distance from the contour line to the core position
defines the angular size.  To avoid measuring distances to spurious
peaks above the 2\% level, we confined the contour to be within the
same rectangular area that was previously determined to completely
enclose the source.

We define integrated core flux densities as the pixel value at the
core component, where the pixel values have units of Jy beam$^{-1}$
and we assume the core itself to be unresolved.  Due to complex jet
structure, rather than sum the pixel values for the jet components we
define the integrated jet flux as the difference between the
integrated total source flux and the core flux.

Due to leakage between orthogonal polarization feeds in the antennas,
we define {\it detected}, polarized cores or jet components as only
those components in which the polarized intensity is greater than
0.2\% of the peak in the total intensity image ({\it p} $>$ 0.002
I$_{\rm peak}$).  We require detected polarizations to be above the
3$\sigma$ level, where $\sigma$ is described by the RMS noise in the
Stokes Q and U maps by the relation $\sigma=(0.21\,\sigma_{U}^{2} +
0.21\,\sigma_{Q}^{2})^{1/2}$ according to Rayleigh statistics.  
In the case that the core has undetected
polarization, we quote {\it m}$_{\rm core}$ as an upper limit equal to
the larger of the two detection restrictions ({\it m}$_{\rm core}
\leq$ max(0.002 I$_{\rm peak}$, 3$\sigma$)).  When studying
correlations between jet axis angle and polarization angle, we
restrict our analysis to those components with $\sigma_{\chi}<20\dg$,
where $\sigma_{\chi}$ is determined by the noise image created by the
AIPS task COMB.

\section{Results}\label{Results and Discussion}

Measured properties for each source are presented in Table 1.
Additional information, including measured jet properties, is
available on the world wide web\footnote{\tt
www.aoc.nrao.edu/$\sim$gtaylor/cjftab.text}.  Images are available
upon request to the authors.  Below we present a statistical analysis
of source properties.

\subsection{Core Fraction}

\begin{figure} 
\epsscale{0.8} 
\plotone{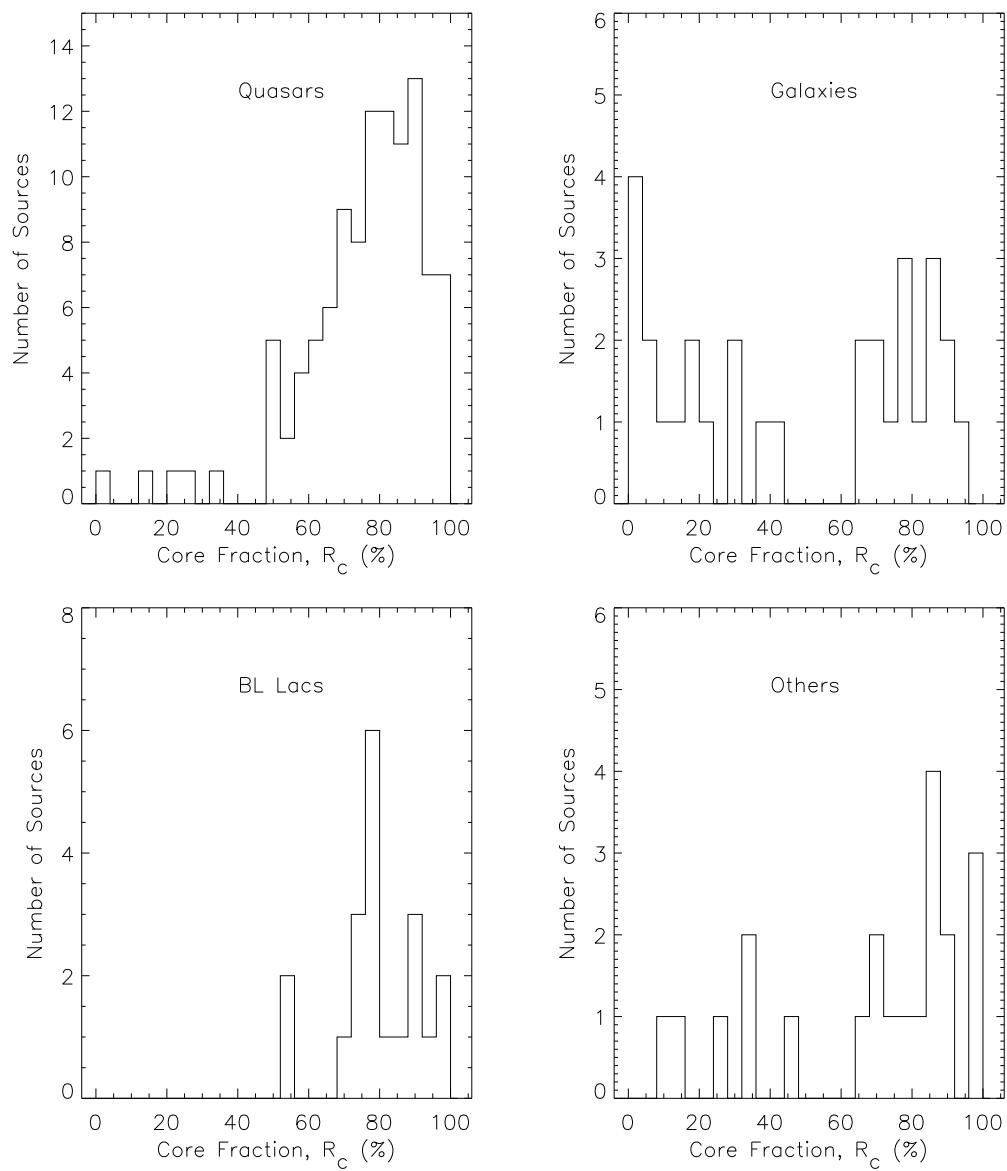}
\caption{Distribution of core fractions, R$_{c}$, in quasars,
galaxies, BL Lacertae objects and others, given as a percentage.
R$_{c}$ is the Stokes I flux density in the core divided by the
integrated Stokes I flux density in the source, where we have
calculated the core flux as described in \S~\ref{Defs}.  The width of
each bin is 4\%.  All sources are shown.}
\label{corefract}
\end{figure}

From Fig.~\ref{corefract} it is clear that the core fractions of
galaxies come from a different parent distribution than that of
quasars or BL Lacertae objects.  While quasars and BL Lacs generally
exhibit a core fraction (defined $R_{\rm c} = S_{\rm c}/S_{\rm
total}$) $R_{\rm c}\approx$80\%, the 30 galaxies in our sample show no
tendency toward either high or low core fraction.  The tendency towards
higher core fraction in quasar and BL Lacs is expected from unified
schemes since the core fraction is a strong function of orientation.   The
Kolmogorov-Smirnov (K-S) test disproves the null hypothesis that the
distribution of quasars ($d_{\rm Q}$) and the distribution of BL Lacs
($d_{\rm BL}$) are the same as the distribution of galaxy core
fractions ($d_{\rm G}$).  When comparing $d_{\rm Q}$ to $d_{\rm G}$
the K-S test outputs the maximum value of the absolute difference
between the two cumulative distribution functions (c.d.f.'s), $D$=0.45
with a probability, $p$, that quasars and galaxies are drawn from the
same population of $7.7 \times 10^{-5}$.  Similarly, when comparing
$d_{\rm BL}$ to $d_{\rm G}$, we find a probability of $2.9 \times
10^{-3}$.  Comparing $d_{\rm BL}$ to $d_{\rm Q}$ we find $p$=0.41,
indicating that these samples could come from the same parent
distribution.  Since the cores and jets generally have
different spectral indices, the core fraction will depend weekly on
redshift.  As an example, a source with $R_{\rm c}$ = 0.75 at a redshift of
zero would shift to $R_c$ = 0.83 at z=1 and $R_{\rm c}$ = 0.87 at z=2,
assuming a flat spectrum core and steep ($\alpha = -0.7$) jets.  In
Fig.~\ref{corefract} we have not attempted any redshift correction
so this could account for the minor difference between the shape of
the distribution for the quasars and BL Lac objects.

The distribution of $R_{\rm c}$ in Fig.~\ref{corefract} for the
galaxies appears somewhat bimodal.  Of the 15 sources with $R_{\rm c}
< 0.5$ more than half (9/15) are CSOs.  No obvious properties are
identified with the high $R_{\rm c}$ population, although our knowledge
of these sources is not complete.

One might be concerned that this comparison uses a sample of quasars
four to five times as large as the samples of BL Lacs and radio
galaxies, respectively. This should not be a factor if we consider the
effective number of data points $N_{e}$ as defined by
\citet{numrec}. $N_{e}$ is defined as

\begin{equation}
N_{e} = \frac{N_{1}N_{2}}{N_{1} + N_{2}}
\end{equation}

\noindent where $N_{i}$ are the number of data points in the
respective sets.  As long as $N_{e}$ is greater than or equal to four
the probability estimate should be accurate. For the quasar$-$galaxy
comparison $N_{e} = 23.4$, quasar$-$BL Lac $N_{e} = 16.8$, and for BL
Lac$-$galaxy $N_{e} = 12.0$.

\subsection{Core and Jet Fractional Polarizations}

\begin{figure} 
\epsscale{0.8} \plotone{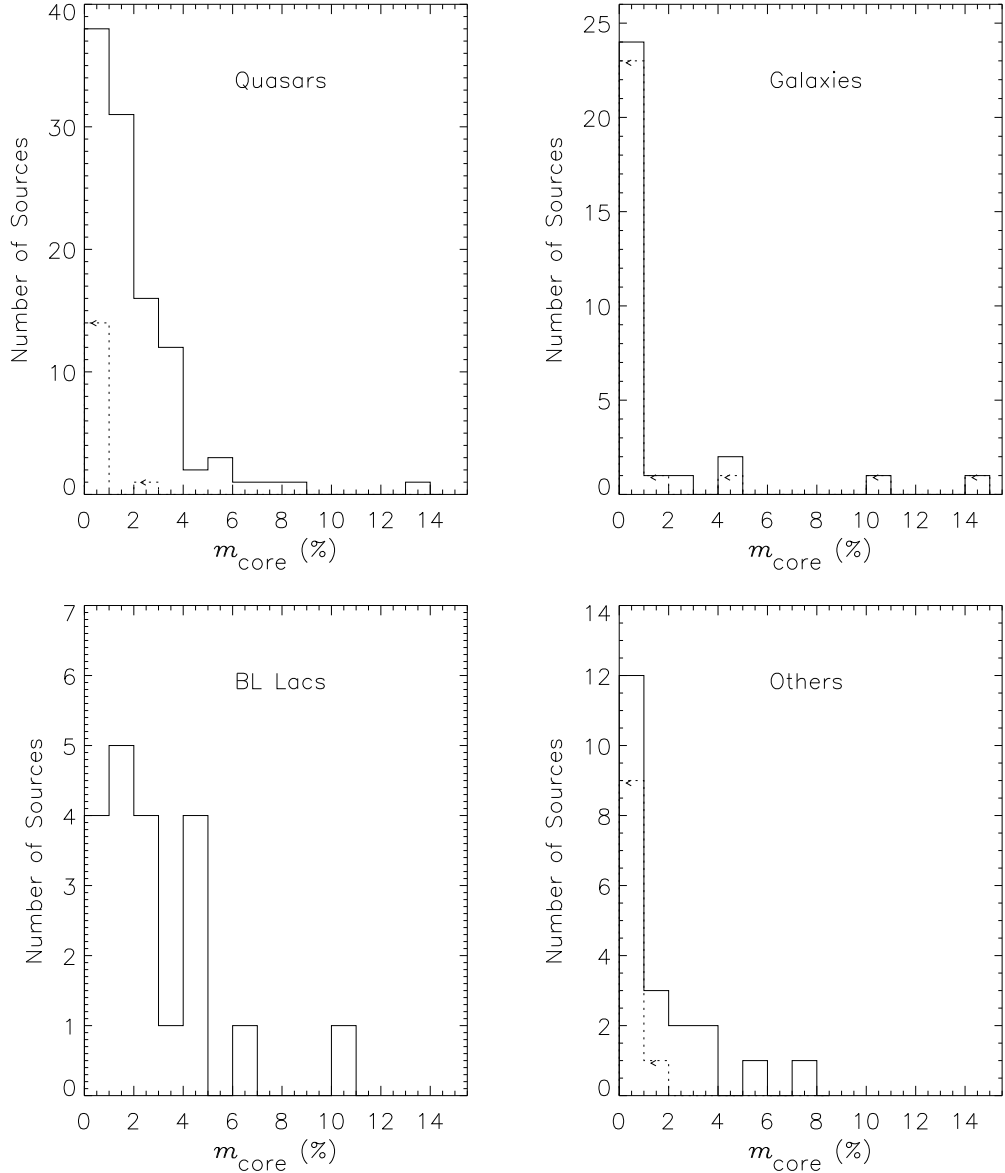}
\caption{Distribution of the fractional polarization ({\it m}) in the
cores of quasars, galaxies, BL Lacertae objects, and others, given as
a percentage.  Upper limits are shown as dashed lines with the $<$
symbol.  The solid line represents the sum of the upper limits and
detections in each bin.  Upper limits were assumed when the polarized
intensity {\it p} $<$ 0.2\% of the peak in the total intensity image,
or when {\it p} $<$ 3$\sigma$, where $\sigma$ is the RMS noise
calculated as described in \S~\ref{Defs}.  The quasar distribution
represents 106 sources with 15 upper limits.  The galaxy distribution
represents 30 sources with 27 upper limits.  The BL Lacertae
distribution represents 20 sources with no upper limits, and the
distribution of other sources represents 21 sources with 10 upper
limits.  The width of each bin is 1\%.}
\label{corepp}
\end{figure}

Fig.~\ref{corepp} shows the distribution of average fractional
polarization in the cores of quasars, galaxies and BL Lac objects.
From this distribution we see that quasars have typical core
polarizations of a few percent, and most (87\%) are detected.
Galaxies are clearly less polarized than quasars on the parsec scale
with only 3 of 30 (10\%) of the sources found to have detectable
polarization in their cores.  No detectable polarized flux was found
from any of the 9 CSOs observed.  This is consistent with results from
a VLBI follow-up survey of CSOs by \citet{peck00} who found less than
1.2 mJy of polarized flux from each of 21 sources observed at 8.4 GHz
with the VLBA.  Generally we find less than 5 mJy of polarized flux
density and fractional polarizations of $<$0.5\%.  For a few CSOs with
weak cores, our upper limits on the fractional polarization are quite
high.  In contrast the BL Lacs appear to have slightly more strongly
polarized cores than quasars (K-S test gives $p$=0.20 that they are
drawn from the same population).

The K-S test only compares detections, and ignores the upper limits.
Survival analysis \citep{fn85} takes into account the upper limits 
in the comparison of two samples. Using the Penn State ASURV 1.2 package 
\citep{lif92} we tested the hypothesis that the BL Lac and quasar 
core fractional polarization results were drawn from the same parent
distribution. Including the 15 upper limit quasar results decreased 
the likelihood that the two samples were drawn from the same parent
distribution. The five two sample tests in ASURV 1.2 give an average
probability of 1.5\% that the null hypothesis is correct. This is 
significantly different from the K-S test result, and we find that 
this result is not affected by eliminating the one rather high upper 
limit quasar core fractional polarization of 2.7\%.

\begin{figure} 
\epsscale{0.8} 
\plotone{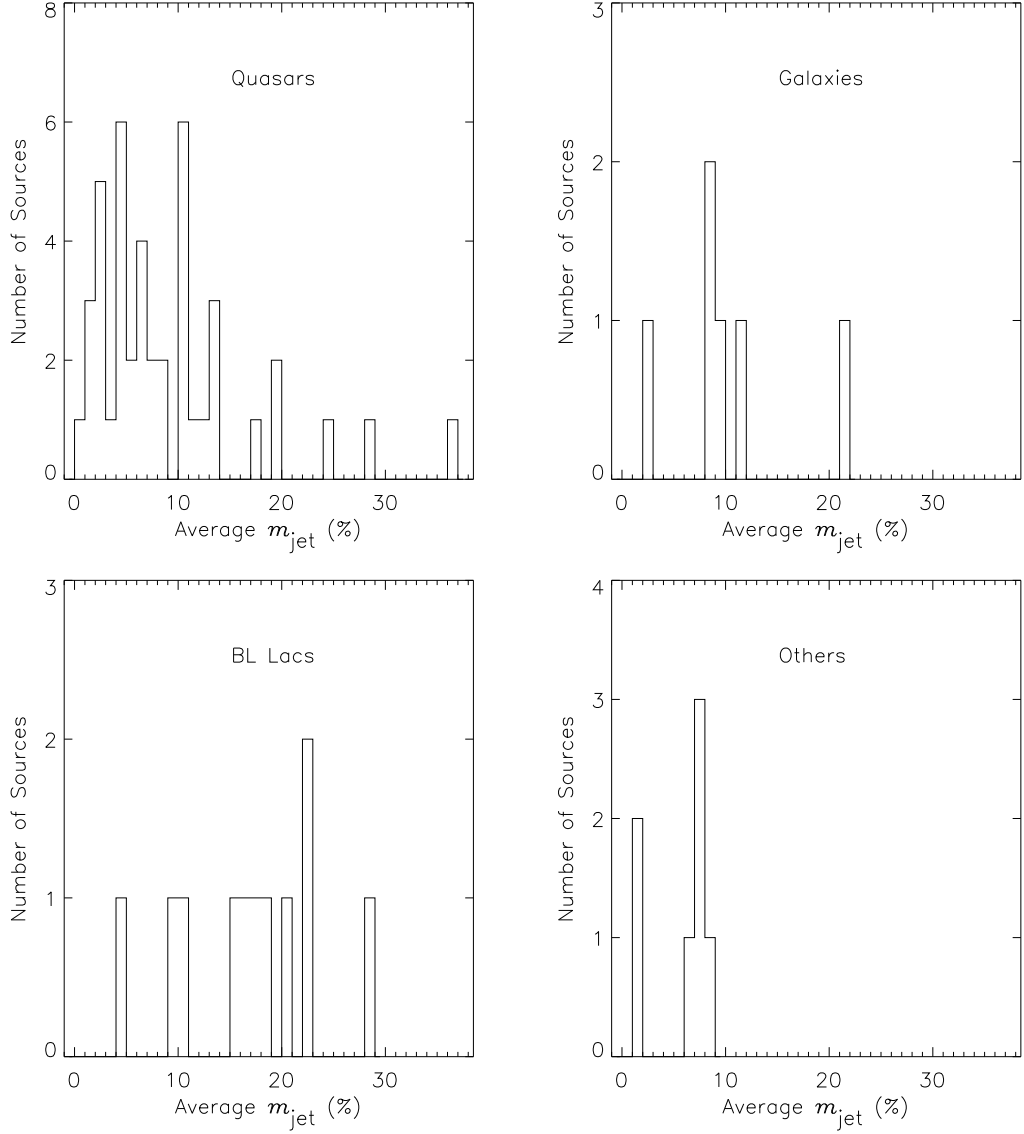}
\caption{Distribution of the average fractional polarization ({\it m})
in the jets of quasars, galaxies, BL Lacertae objects, and others,
given as a percentage.  Only detected jet components have been
averaged.  (See \S~\ref{Defs} for a definition of detected components.)
The distribution of quasars represents 43 of the 106 we have in our
sample.  21 quasars were excluded due to their naked core morphology,
and 42 sources had no jet components with detected polarizations.  The
distribution of galaxies represents 6 sources.  Four were excluded as
they were naked cores, and 20 had no detected jet components.  The
distribution of BL Lacertae objects represents 11 sources.  Three BL
Lacs were naked cores and 6 had no detected jet components.  The
distribution of others represents 7 sources.  6 were naked cores and 8
had undetected jet components.  The width of each bin is 1\%.}
\label{jetpp}
\end{figure}

\begin{figure} 
\epsscale{0.9} 
\plotone{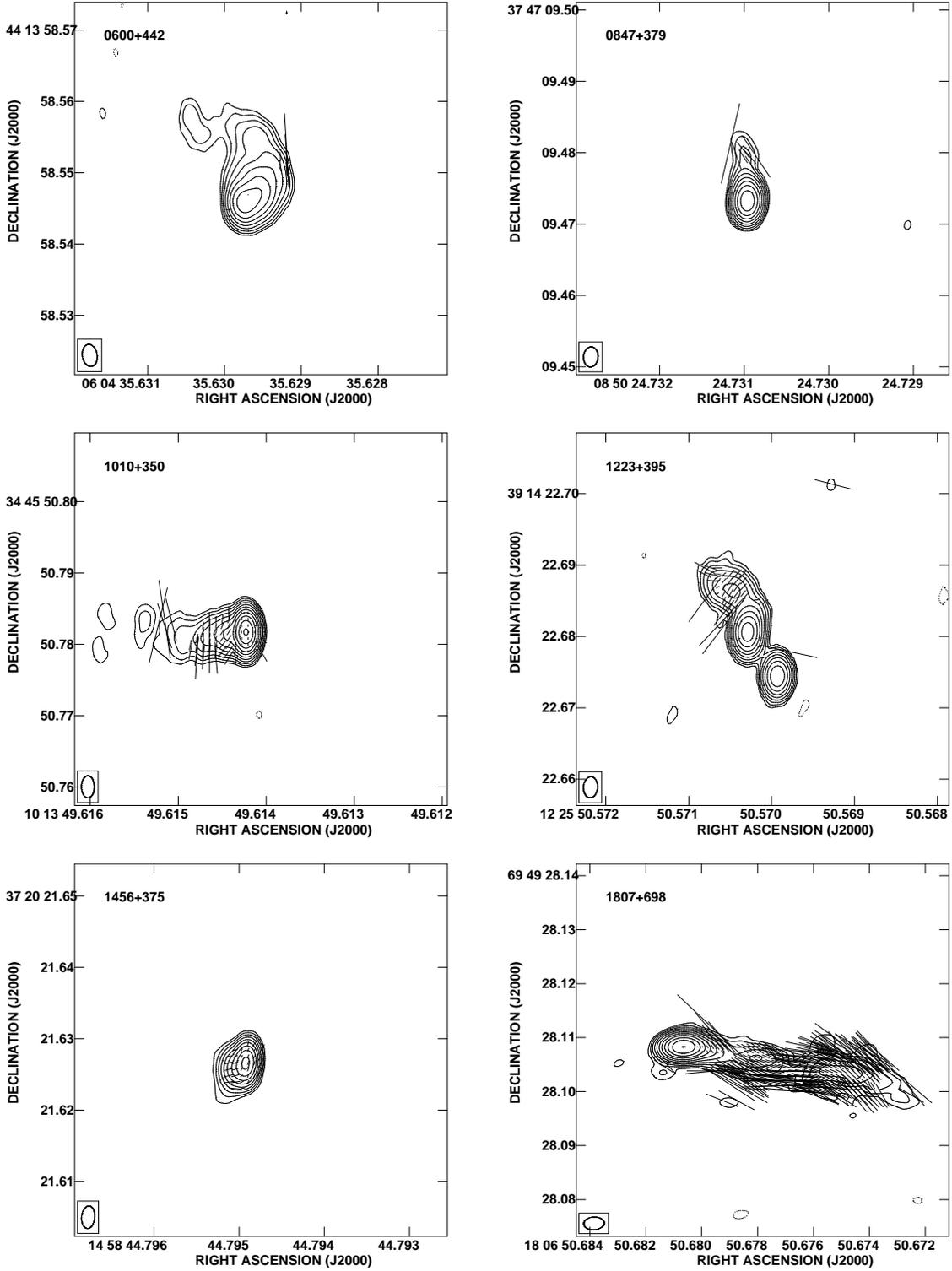}
\caption{A plot of all six radio galaxies within our sample having
some detected polarization.   Contours are drawn
at $-4\sigma_I$, $4\sigma_I$, $8\sigma_I$, $16\sigma_I$, ..., where
$\sigma_I$ is RMS of total intensity given in Table 1.  Polarization
vectors have lengths proportional to fractional polarization.
Blanking of the polarization has been performed on pixels less than 
3$\sigma_p$ in polarization, 4$\sigma_I$ in total intensity, or 0.2\%
in fractional polarization.}
\label{galmos}
\end{figure}

Fig.~\ref{jetpp} shows the same set of distributions, but for the
average fractional polarizations of the jet components.  In this case
the fractional polarization of the BL Lacs is significantly greater
from that of the quasars (p=0.0037 that they are drawn from
the same population).  When the jet polarization is detected, sources
have higher fractional polarization in the jets than in the cores.
Even the galaxies have an average fractional polarization in the jets
of 10\%, compared to typical fractional core polarizations of $<$1\%.
In Fig.~\ref{galmos} we show the images for all six radio 
galaxies with some detected polarization.  The measurements range 
from marginal in 0600+442 and 0847+379 to extremely detailed 
in the bright, one-sided jet of 1807+698.

\subsection{Comparing Alignments Between Jet Axis and EVPA}

Many claims have been made in the literature regarding comparisons of
the orientation of the jet axis ($\theta$) with the EVPA of the jet
($\chi_{\rm jet}$) or core ($\chi_{\rm core}$).  Below we discuss
various tests to look for preferred {\it alignments}.  Ideally these
tests should be made after having removed any Faraday rotation due to
the ISM of our Galaxy, or the environment local to the sources.
Unfortunately, for these single-frequency observations it was not
possible to estimate the Rotation Measure (RM).  If significant RMs
exist towards any of the source components (e.g., as found to be
common in quasar cores by Taylor 1998, 2000) then that will tend to
smear out any intrinsic correlation.

The rotation measure of the cores of BL Lacs are typically 2$-$300
\radm, and those of quasars range from 200 \radm\ to several 1000 
\radm\ \citep{zav03}. At 5 GHz a rotation measure of 277 \radm\ is
enough to cause a turn of one radian. This would be sufficient to
smear out any correlation in EVPA. The RMs in the jets 
of 16 BL Lacs, radio galaxies and quasars, with one exception, 
are all approximately 200 \radm, and thus would also smear out any 
correlation in $\theta - \chi_{\rm jet}$ \citep{zav03}.

\begin{figure} 
\epsscale{0.8} 
\plotone{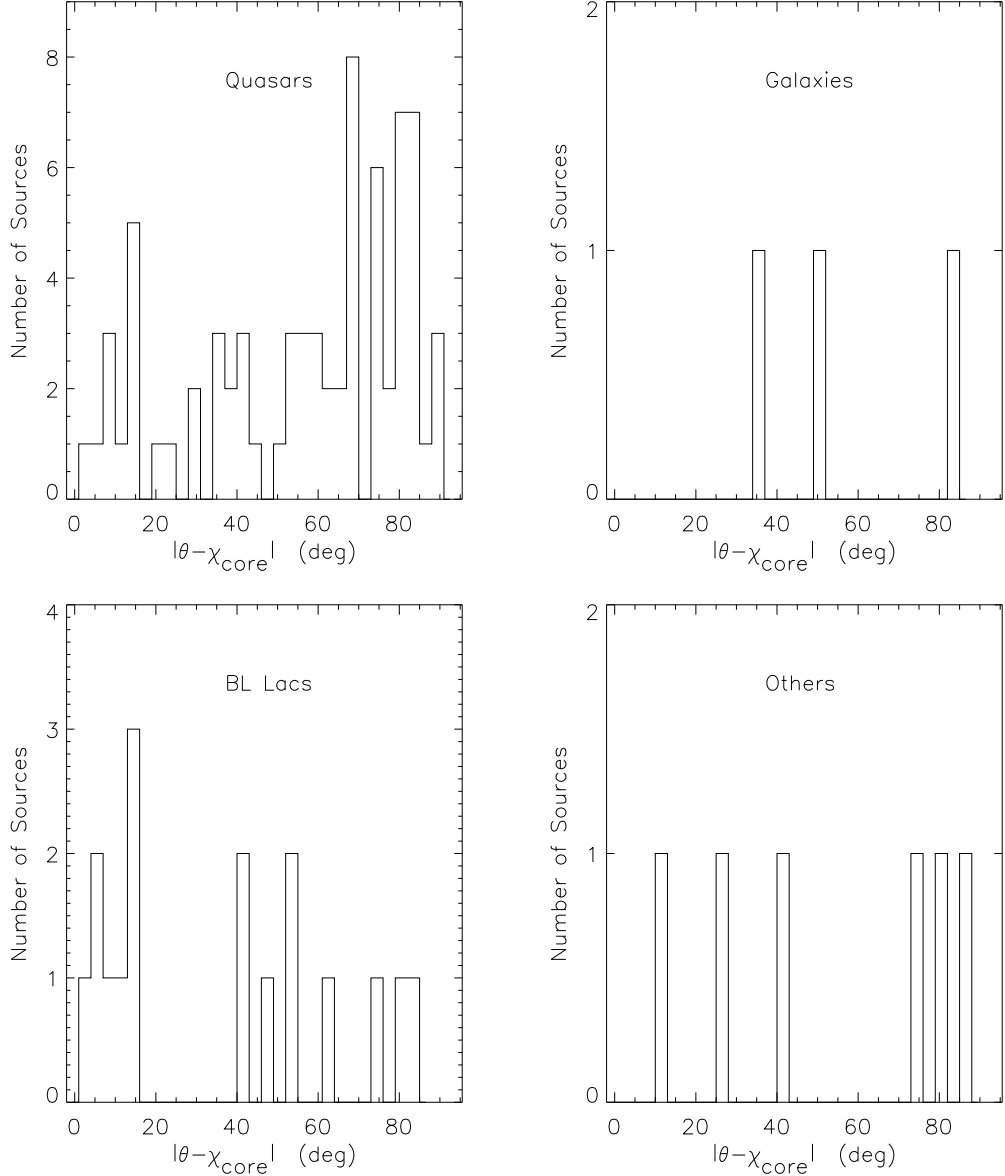}
\caption{Distribution of the difference between the jet axis angle,
$\theta$, and the electric vector position angle at the core,
$\chi_{\rm core}$ given in degrees.  The distribution of quasars
represents 72 sources.  21 quasars were excluded due to their naked
core morphologies, and 13 were excluded as they did not meet the
detection criteria described in \S~\ref{Defs}.  The distribution of
galaxies represents just 3 sources, where 4 galaxies were found to
have naked core morphologies and 23 had undetected cores.  17 BL
Lacertae objects are shown.  Three naked core BL Lacs were excluded.
The distribution of others represents 6 sources.  9 had undetected
core polarizations, and 6 were naked cores.  The width of each bin is 3\deg.}
\label{corejetaxis}
\end{figure}

In Fig.~\ref{corejetaxis} we present the distribution of $\theta -
\chi_{\rm core}$ for quasars, galaxies and BL Lac objects.  For the
quasars we see that the distribution is not quite flat, and hints at
an excess of sources near 80 degrees.  These nearly misaligned EVPAs
imply a magnetic field nearly perpendicular to the jet axis assuming
optically thick synchrotron emission.  The K-S test indicates only a
0.3\% chance that the distribution is the same as a flat distribution,
suggesting the possibility of an intrinsic correlation between EVPA
and $\theta$.  There is a somewhat greater probability (3.7\%) that
the quasars and BL Lacs are drawn from the same population.  However,
the BL Lac distribution itself is not significantly different from a
flat distribution (17\% chance that it is the same as flat). 

\begin{figure} 
\epsscale{0.8} 
\plotone{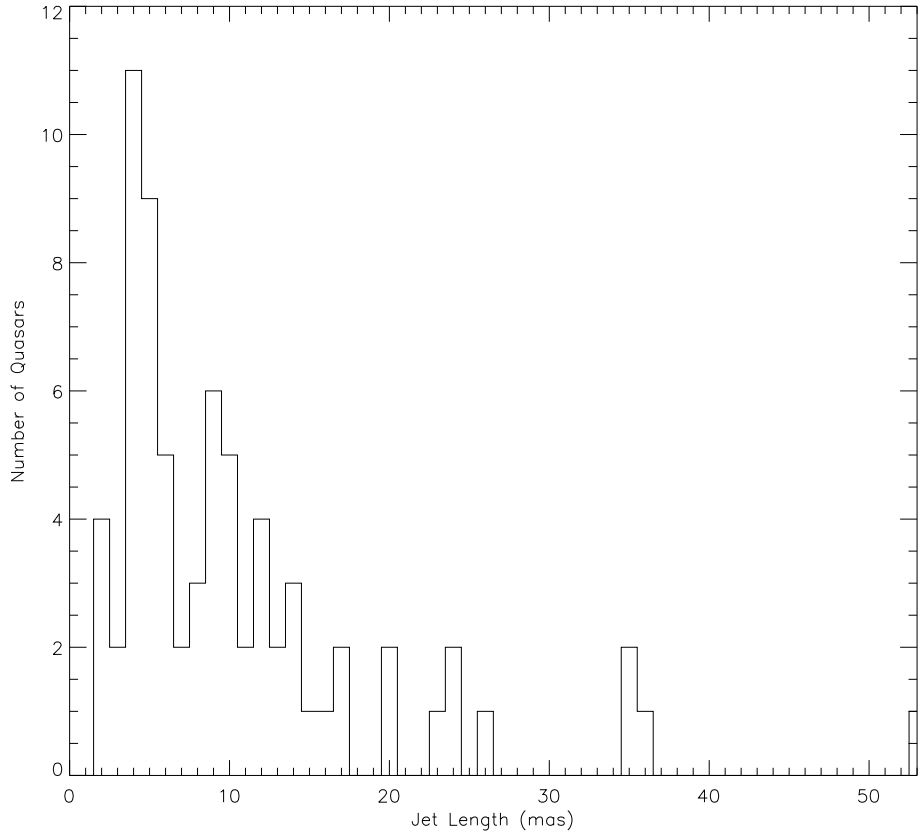}
\caption{Distribution of uncorrected jet lengths for the 72 quasars
shown in Fig.~\ref{corejetaxis}.  The width of each bin is 1 mas.}
\label{6mas}
\end{figure}

\begin{figure} 
\epsscale{0.8} 
\plotone{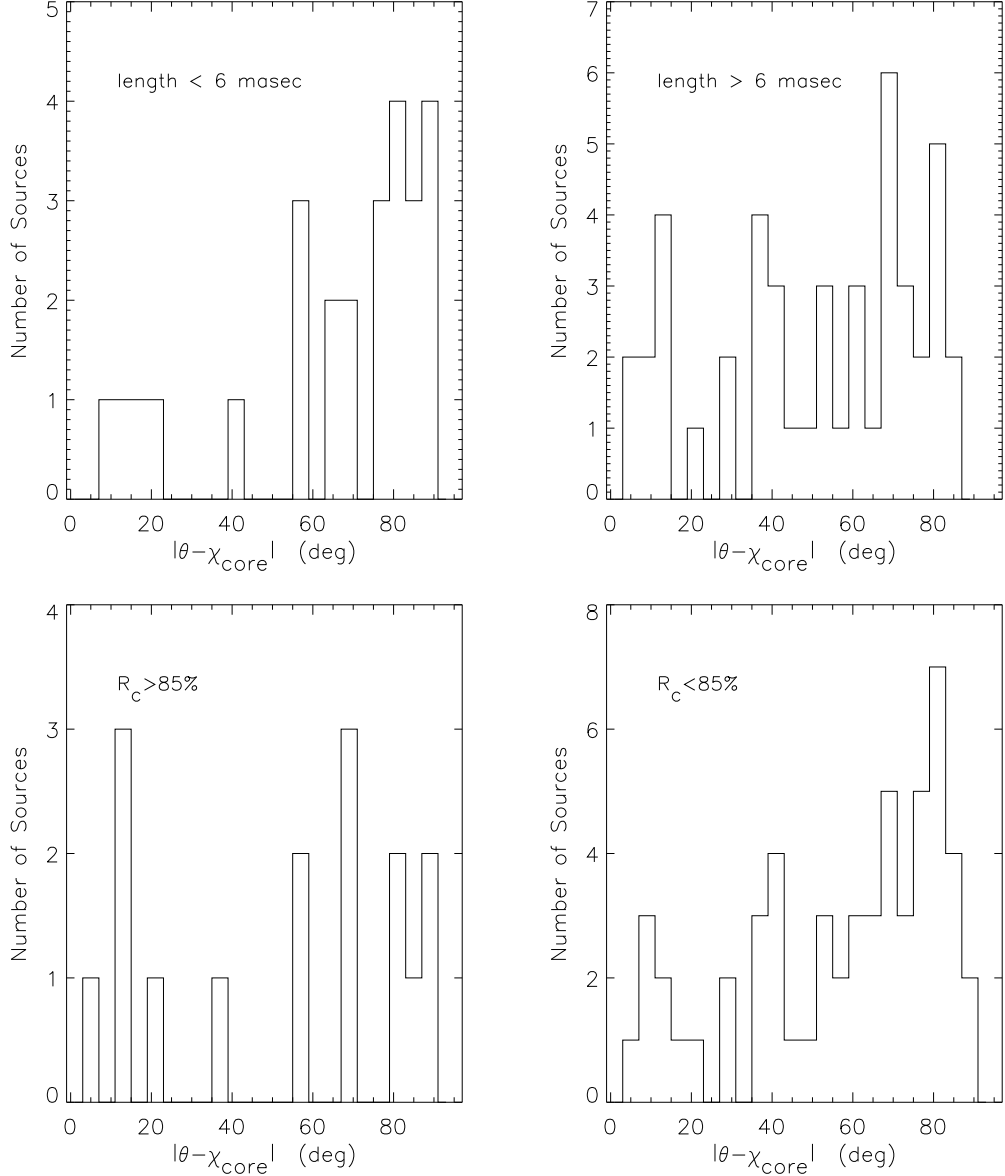}
\caption{Distribution of the difference between the jet axis angle,
$\theta$, and the electric vector position angle at the core,
$\chi_{\rm core}$, for quasars.  Top left: $\theta-\chi_{\rm core}$ relation
for those quasars with uncorrected lengths of less than 6 mas.  26
quasars are shown.  Top right: relation for quasars with lengths
greater than 6 mas.  46 quasars are shown.  Bottom left:
$\theta-\chi_{\rm core}$ relation for those quasars with core fractions
greater than 0.85.  16 quasars are shown.  Bottom right: relation for
quasars with core fractions less than 0.85.  56 quasars are shown.
Here we measure length as the angular distance from the core to the
farthest jet component, irrespective of jet bend.  The width of each
bin is 4\deg.}
\label{corejetaxislength}
\end{figure}

To further investigate the nature of the misaligned core EVPAs we
divided the quasars into those with short (length $<$ 6 mas) jets and
long (length $>$ 6 mas) jets.  From Fig.~\ref{6mas} we see that a jet
length of 6 mas divides the quasars into two groups of roughly equal
sizes since there are jet length peaks at 4 and 9 mas.  We used
ASURV's correlation tests \citep{ifn86} to examine the correlation
that appears in Fig.~\ref{corejetaxislength} for the quasars with
short jets.  Although it appears that short jets are correlated with
misaligned core EVPAs, a Kendall's tau test gives a less than 1 sigma
significance of a correlation. Additionally, plots of $|\theta -
\chi_{core}|$ versus jet length show no obvious correlation,
regardless of jet length. Although there is no evidence for a
correlation of quasar core EVPA alignment with jet length the K$-$S
test of quasars versus a uniform distribution indicates that quasars
have a preference for misaligned core EVPAs. As one can see from
Fig.~\ref{corejetaxislength} the cores associated with short jets
appear more misaligned (probability of a flat distribution is 0.009\%,
probability that they have the same distribution as sources with long
jets is 1.7\%), with a clear excess near 90 degrees.  We also divided
the quasars into those with a high ($R_c > 85$\%) or low ($R_c <
85$\%) core fraction.  Core fraction seems much less important as a
discriminator of the misaligned core EVPA population.

\begin{figure} 
\epsscale{0.8}
\plotone{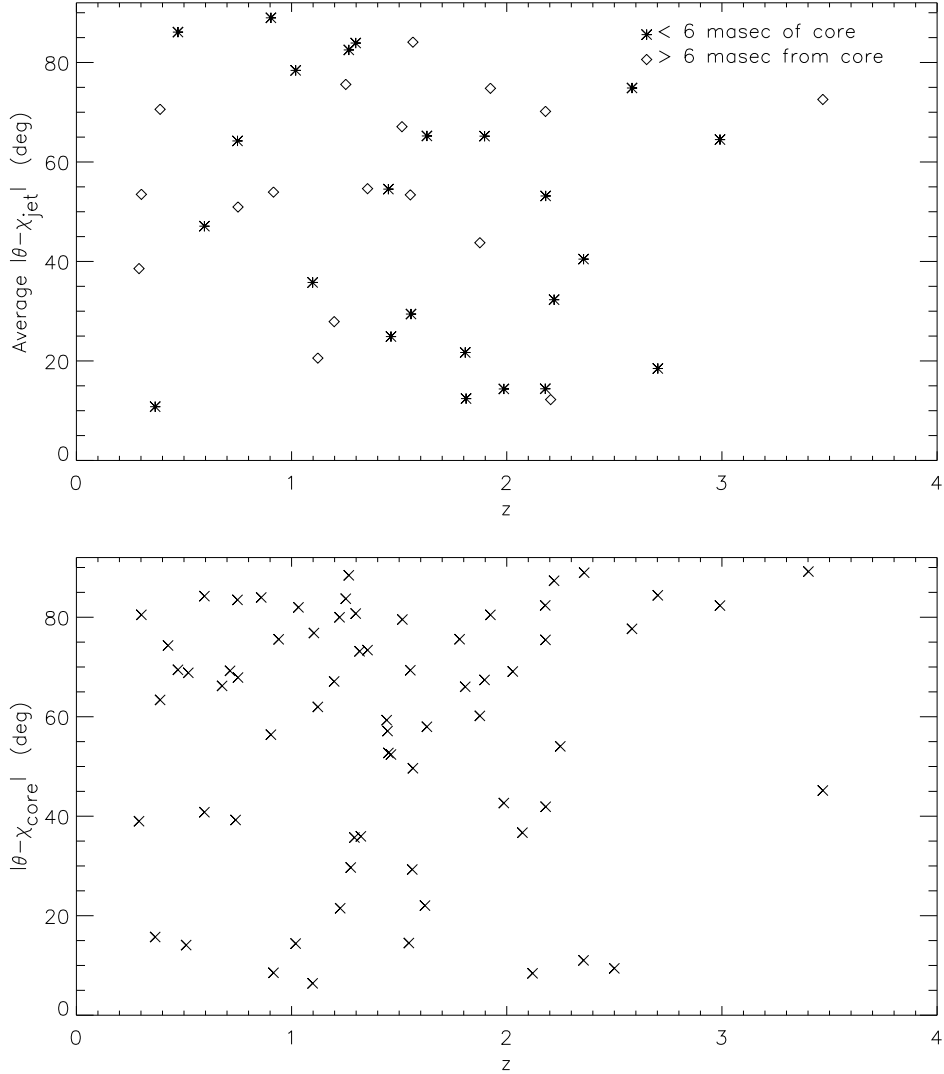}
\caption{Top: Plot of the average difference between the jet axis
angle, $\theta$, and the electric vector position angle in the
detected jet components of quasars, $\chi_{\rm jet}$, as a function of
redshift.  We have averaged the jet components within 6 mas of the
core separately from those farther than 6 mas from the core, so that
any quasar with detected jet components both farther and nearer than 6
mas from the core is represented twice.  24 quasars with detected
jet components within 6 mas of the core are shown, and 17 quasars with
detected jet components farther than 6 mas of the core are shown.
Bottom: Plot of $|\theta-\chi_{\rm core}|$ for those quasars with
detected core polarizations as a function of redshift.  68 quasars are
shown.}
\label{zjetaxisstuff}
\end{figure}

One can also look for any dependence on alignment with redshift.  In
Fig.~\ref{zjetaxisstuff} we plot the alignments of the jet and core
EVPAs with jet axis against redshift.  No trends with redshift are
apparent for either the jets or the cores.  Correlation tests
performed using the ASURV package verified that no correlation of
either core or jet misalignment with redshift is present.  This was
true for fractional polarization in the cores and jets as well.

\begin{figure} 
\epsscale{0.85} 
\plotone{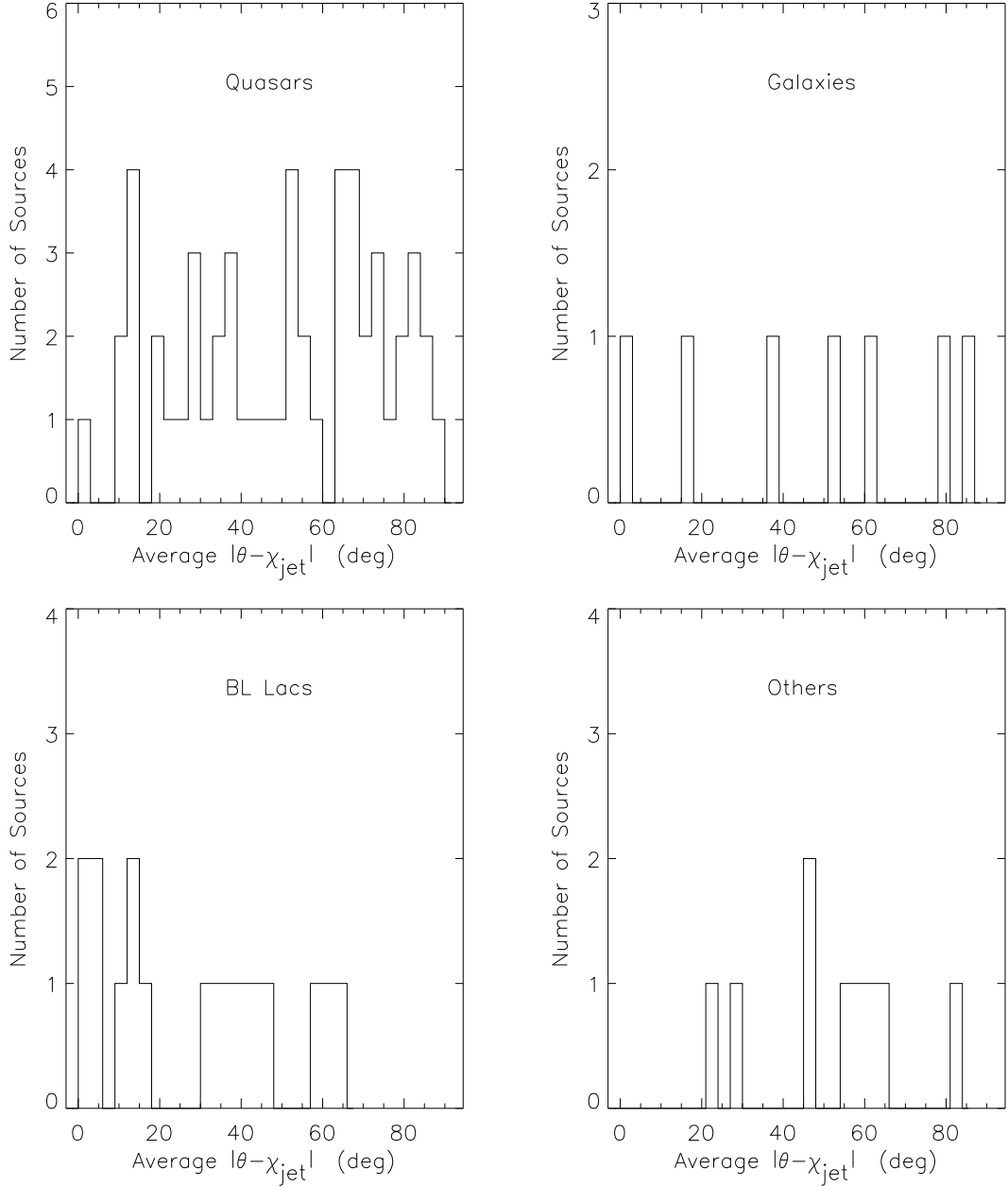}
\caption{Distribution of the average difference between the jet axis
angle, $\theta$, and the electric vector position angle of all
detected jet components.  The
distribution of quasars represents 73 measurements.  The distribution of galaxies represents 7
sources.  The
distribution of BL Lacertae objects represents 17 measurements.  The distribution of others includes 9 sources.  The width of each bin is 3\deg.}
\label{jetjetaxisall}
\end{figure}

In Fig.~\ref{jetjetaxisall} we present the distribution of $\theta -
\chi_{\rm jet}$ for quasars, galaxies and BL Lac objects for the jet
components both near ($<$ 6 mas) and far ($>$ 6 mas) from the core.
Contrary to claims in the literature \citep{caw93,hom02} that quasars
have preferred alignments with the magnetic field oriented parallel to
the jet axis, we see no such correlation.  (The probability of being
drawn from a flat distribution is 34\%.)  We have also tried examining
separately the near ($<$ 6 mas) and far ($>$ 6 mas) jet components and
reach the same conclusions. Cawthorne et al. did apply a 
rotation measure correction, which 
might reveal a correlation our non-RM corrected results hide. 
However, their RMs were derived from the VLA, which does not 
correctly sample the parsec-scale RM of these AGN. Additionally, 
the RM corrections were primarily (12/17) less than 50 \radm, 
which amounts to a correction of 10\deg\ or less at 5 GHz. 
As shown in \citet{zav03} these corrections were too small 
to account for the known parsec-scale RMs. Homan et al. 
did not apply a RM correction, but their observations were
at 15 and 22 GHz. Their results are thus less susceptible 
to any rotation measure smearing. 

The distribution for BL Lac objects (Fig.~\ref{jetjetaxisall}) appears
to have a lack of sources with EVPA perpendicular to the jet.  From
the K-S test we find an 8.0\% chance of this population being drawn
from a uniform $\theta - \chi_{\rm jet}$ distribution.  A bimodal
distribution was found by \citet{gab00} with peaks near 0$\dg$ and
90$\dg$.  Gabuzda et al. performed a similar analysis on a 1 Jy flux
limited sample of BL Lac objects observed with VLBI polarimetry at 5
GHz.  This is shown in Fig. 12 of \citet{gab00}. A K-S test of the
data in their Fig. 12 yields a 1.1\% probability that the data are
drawn from a uniform distribution.  With assistance from Gabuzda
(2002, private communication) we combined our BL Lac $\theta -
\chi_{\rm jet}$ data with the 35 measurements from \citet{gab00} and
tested this data against a uniform distribution in a K-S test. The
combination of these two samples results in a 0.7\% probability that
the parent distribution is uniform.  However, with just the 17 data
points from our sample it is difficult to draw a conclusion on the
nature of the underlying distribution, and we cannot hope to isolate
what type of source gives rise to the slight excess of sources with
EVPA at small angles to the jet axis.

\begin{figure} 
\epsscale{0.9} 
\plotone{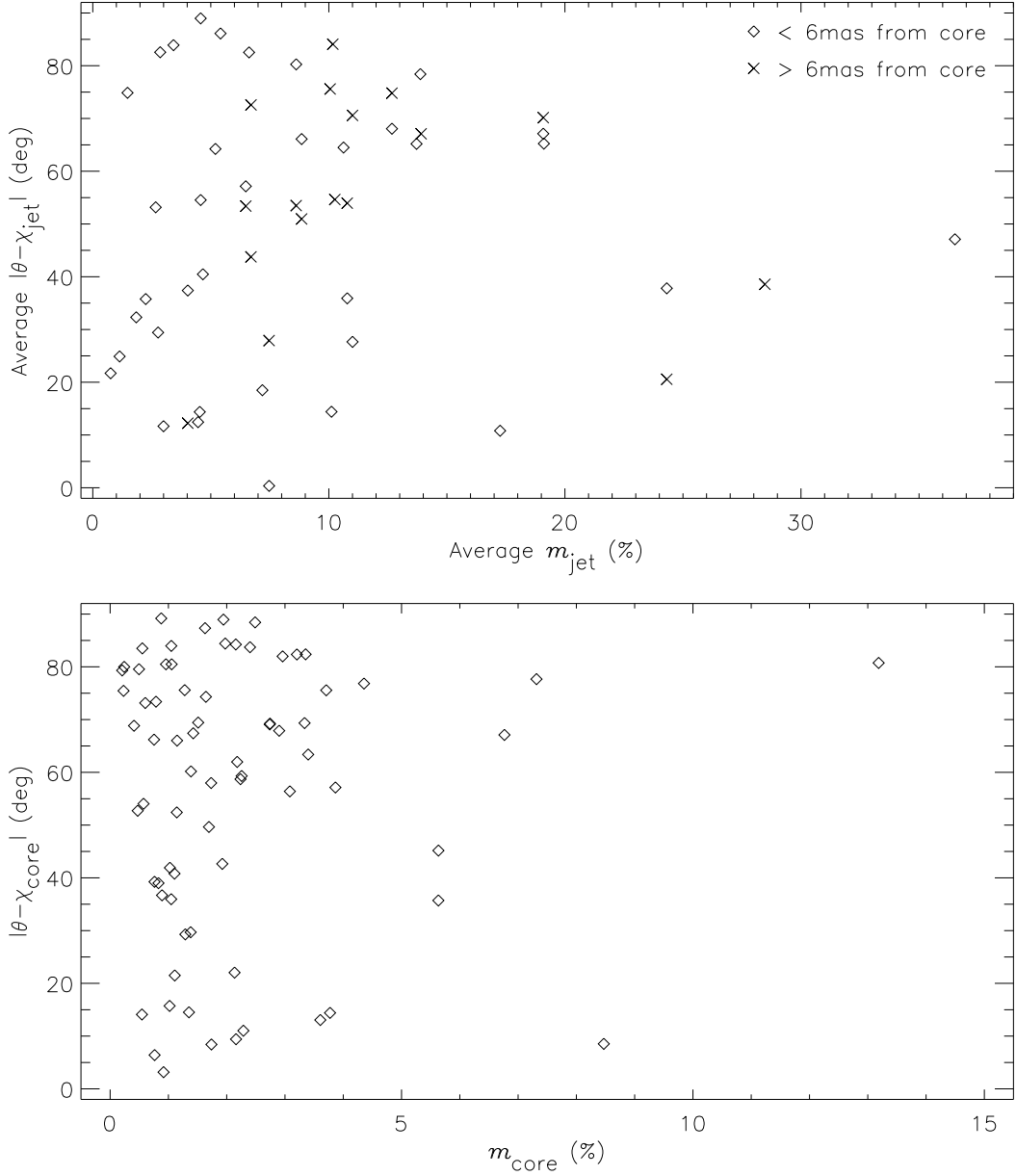}
\caption{Top: Plot of the average difference between the jet axis
angle, $\theta$, and the electric vector position angle of all
detected jet components nearer ($\diamond$) and farther ($\times$)
than 6 mas from the core as a function of the average fractional
polarization of the quasar jet.  Only those quasars with detected jet
polarizations are shown.  36 quasars are represented with the
$\diamond$ symbol and 17 are represented with the $\times$ symbol.
Bottom: Plot of the difference between the jet axis angle and
the electric vector position angle in detected quasar cores as a function of core
fractional polarizations.}
\label{jetppjetjetaxis}
\end{figure}

In a 43 GHz polarimetry survey of flat-spectrum sources from the
Pearson-Readhead survey \citep{pr88}, \citet{list01} found that the
most strongly polarized quasar cores display EVPAs that are aligned
with the jet axis.  In Fig.~\ref{jetppjetjetaxis} we examine the
relationship between alignment and fractional polarization.  No
preferred alignment for the more polarized sources, as what
\citet{list01} found, is evident in our sample.  Also the preferred
core EVPA orientation in our sample is for EVPAs perpendicular to the
jet (Fig.~\ref{corejetaxis}).  These two claims can be reconciled with
the same projected magnetic field orientation if the cores are
optically thin at 43 GHz and optically thick at 5 GHz.

Using the ASURV bivariate tests \citep{ifn86} we examined the 
hypothesis that there is a correlation between the core fractional 
polarization and the core EVPA alignment. The Kendall's tau test 
gives a probability of 82.7\% and the Cox regression a probability 
of 96.9\% that a correlation does not exist. Although there appears
to be a slight excess of sources with misaligned core EVPAs
at fractional polarizations greater than 2.5\%, this is not born
out by the bivariate tests. The Cox regression gives a probability 
of 94.7\% and the Kendall's tau test a 43.6\% probability that no 
correlation is present. The disparity probably results from small 
number statistics as only 20 sources have m$_{\rm core} >$ 2.5\%.


\subsection{Fractional Polarization and Jet Length}
  
In Fig.~\ref{lengthstuff} the core and average jet fractional
polarization is plotted against the projected jet length.  The
fractional jet polarization seems fairly smoothly distributed, but the
fractional core polarization is anti-correlated with jet length.
Sources with short jets seem to have higher core polarizations and
there is a distinct lack of strongly polarized cores in sources with
long jets.  The censoring present in the lower panel of 
Fig.~\ref{lengthstuff} led us to use survival analysis to test for 
a correlation for core EVPA alignment with jet length. Both the 
Kendall's tau and Spearman's rho tests give a weak (1.6 $\sigma$) 
support to the anti-correlation which seems to be present. 
A possible explanation for this trend is that sources with
long jets are viewed at angles farther from the line-of-sight.
Assuming the standard obscuring torus model, the angle to the core
then traverses a higher-density region with magnetic fields causing
Faraday depolarization.  This interpretation is also consistent with
the finding that the cores of BL Lacs (thought to be viewed at very
small angles to the line-of-sight \citep{ant93}) are somewhat more strongly
polarized than quasar cores.

\section{Cosmology Using the Size$-$Redshift Relation}

The $\theta-z$ relation for compact radio sources has been studied in
the past to place limits on $q_{\circ}$.  \citet{kel93} used a sample
of 82 core-jet sources selected from the literature to find that the
deceleration parameter is roughly 0.5, without appealing to source
evolution.  \citet{gur99} did a similar analysis using 330 5 GHz
compact sources taken from the literature, finding that
$q_{\circ}=0.21 \pm 0.30$ if no linear size$-$luminosity, $-$redshift
or $-$spectral index dependences are assumed.  \citet{gur99} applied a
regression model to 145 of the 330 sources with additional
restrictions on luminosity and spectral index.  With this subset of
sources a deceleration parameter of $q_{\circ}=0.33 \pm 0.11$ was
found.  All of these results are consistent with Friedmann cosmologies
with $0 \leq q_{\circ} \leq 0.5$ and $\Lambda$=0.

Figure~\ref{kellermann} shows the $\theta-$z relationship for 103
core-jet sources from our complete sample and mirrors Fig. 1 in
\citet{kel93} and Fig. 5 in \citet{gur99}.  We measured angular sizes
similarly to how Kellermann and Gurvits did, accounting for the
decrease in brightness with redshift as described in \S~\ref{Defs}.
CSOs have been excluded due to the proposed relation between their age
and angular size, and BL Lacertae objects have been excluded because
of their proposed preferred inclination angles and to be consistent
with the methods of \citet{kel93}. The redshifts range from 0.0172
$\leq z \leq$ 3.469.  With the exception of the slope between the
first two redshift bins, our plot shows no clear dependence between
angular size and redshift.  Binning the data in redshift space rather
than restricting each bin to equal numbers of sources does not greatly
alter the plot.  We found that plotting median values rather than mean
values also does not change the plot's shape considerably.  Thus
contrary to previous claims our complete, flux-density limited sample,
although not inconsistent with Friedmann cosmologies, cannot
definitively rule out a steady state universe or place any limits on
$q_{\circ}$.  The few sources with higher angular sizes and reasonable
error bars clustered around z=1.1 are, however, suggestive of a
general increase of angular size with redshift after z=1, which
supports inflationary cosmologies.

Although these compact radio sources with smaller redshifts and
younger ages are not as subject to evolutionary effects, and although
the physical parameters of their central engines are thought to
outweigh any effects due to possible variations of intergalactic
medium with redshift, we find that these objects nevertheless cannot
be used to place restrictions on the deceleration parameter and the
geometry of our universe.  The large error bars in
Fig.~\ref{kellermann} prove these sources to be poor standard rods,
possibly because of their rapid evolution compared to the Hubble time.
It is possible that strong Doppler favoritism is skewing our results
so that more objects with small inclinations to the line-of-sight and
therefore greater, Doppler-boosted luminosities and smaller angular
sizes are getting into the sample without being excluded by the flux
density requirement.  Our neglect of Doppler boosting may also affect
our measurement of angular size since we have defined object size by
the contour line at 2\% of the peak in the total intensity map.
Angular size might also be better measured taking into account jet
bend, rather than assuming a one-dimensional jet in all cases.
Lastly, it is possible that with a multi-frequency study using an
angular size$-$spectral index relation, as is shown in Fig. 7 of
\citet{gur99}, the $\theta-z$ correlation may become tighter.

\subsection{{Bent-Jet Sources and Other Interesting Properties}}\label{Bend}

Although we did not do an exhaustive statistical analysis of the
correlation between jet bending and other source properties, we do
note two basic categories into which most bent sources fall.  One
class of bent jets typically shows little or no polarized intensity in
the jet, while there may be polarized intensity at the source core.
Some examples of this are 0600+442, 0700+470, 0707+476, 0831+557, and
0843+575.  The second and larger class of bent jets shows a tendency
for the electric field to follow the curve of the bending axis so that
the electric vector position angles are usually, but not always,
perpendicular to the bending axis of the jet.  Figure~\ref{bentmos}
shows 0133+479, 0627+532, 1151+408, 1459+480, 1619+491, and 2351+456
which are all typical sources in this second category.  A nice example of an
$\bf{E}$-field bending with and parallel to the jet axis is the
quasar 0627+532.

\section{Conclusions and Future Work}\label{Conclusions}

\begin{figure} 
\epsscale{0.85} 
\plotone{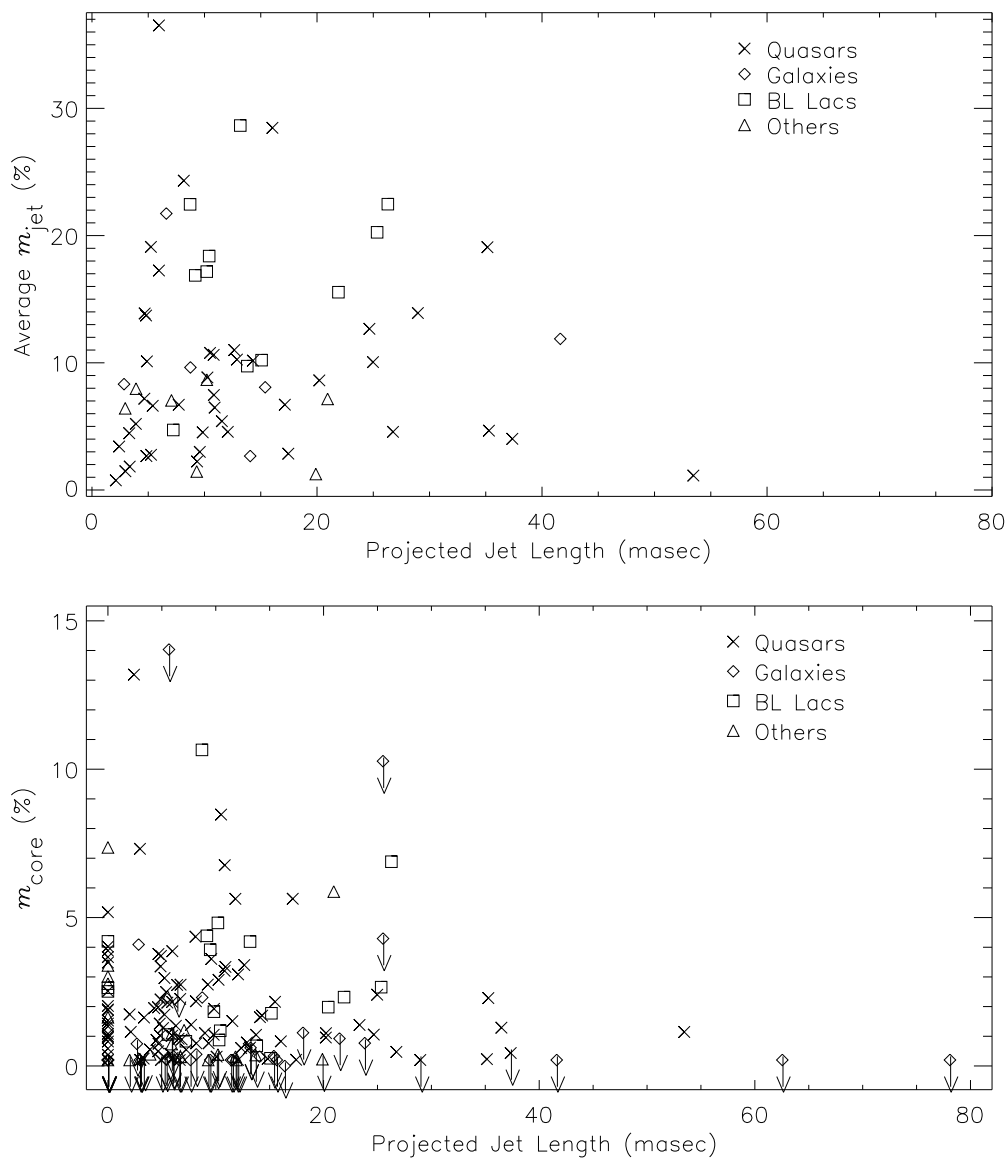}
\caption{Plots of the average fractional polarizations of jet
components (top) and the fractional core polarization (bottom)
vs. projected jet length.  The projected jet lengths and upper limits
for the fractional core polarizations are calculated as described in
\S~\ref{Defs}.  The top plot shows 43 quasars, 6 galaxies, 11 BL Lacs
and 7 others.  The bottom plot shows data for 106 quasars, 30
galaxies, 20 BL Lacs and 21 others.}
\label{lengthstuff}
\end{figure}

\begin{figure} 
\epsscale{0.68} 
\plotone{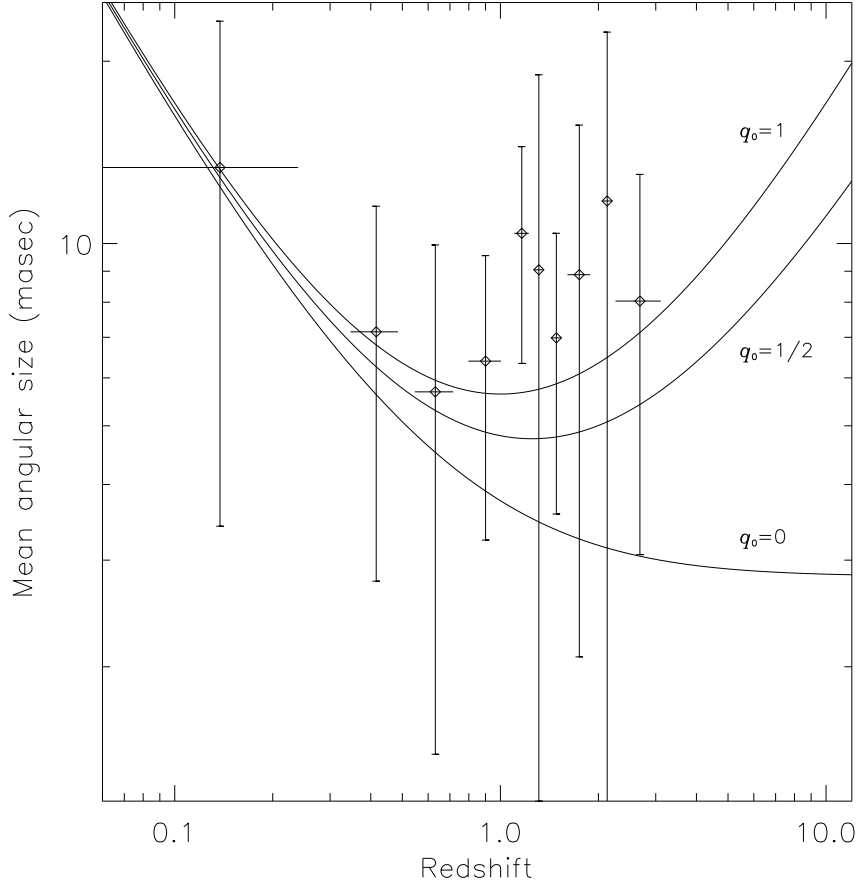}
\caption{Plot of the redshift vs. uncorrected mean angular size in
mas, where the mean angular size is calculated as described in
\S~\ref{Defs}.  Our definition of mean angular size mirrors the
definition given in Kellermann (1993).  Only those sources with
core-jet morphologies and known redshifts are included.  Additionally,
BL Lacertae objects have been omitted to give a total of 103 sources.
Each point represents roughly equal numbers of sources.  (Starting
with the lowest redshift bin, the bins contain 10, 10, 10, 11, 10, 10,
11, 10, 10 and 11 sources, respectively.)  We show 1$\sigma$ error
bars, where $\sigma_x$ is the standard deviation of the redshifts in
each redshift bin, and $\sigma_y$ is the standard deviation of the
angular sizes in each redshift bin.  The solid curves represent the
theoretical dependence of angular size on redshift assuming a
Friedmann cosmology with H$_{\circ}$=50 km s$^{-1}$ Mpc$^{-1}$ and a
source with an angular size of 41 parsecs.}
\label{kellermann}
\end{figure}

\begin{figure} 
\epsscale{0.9} 
\plotone{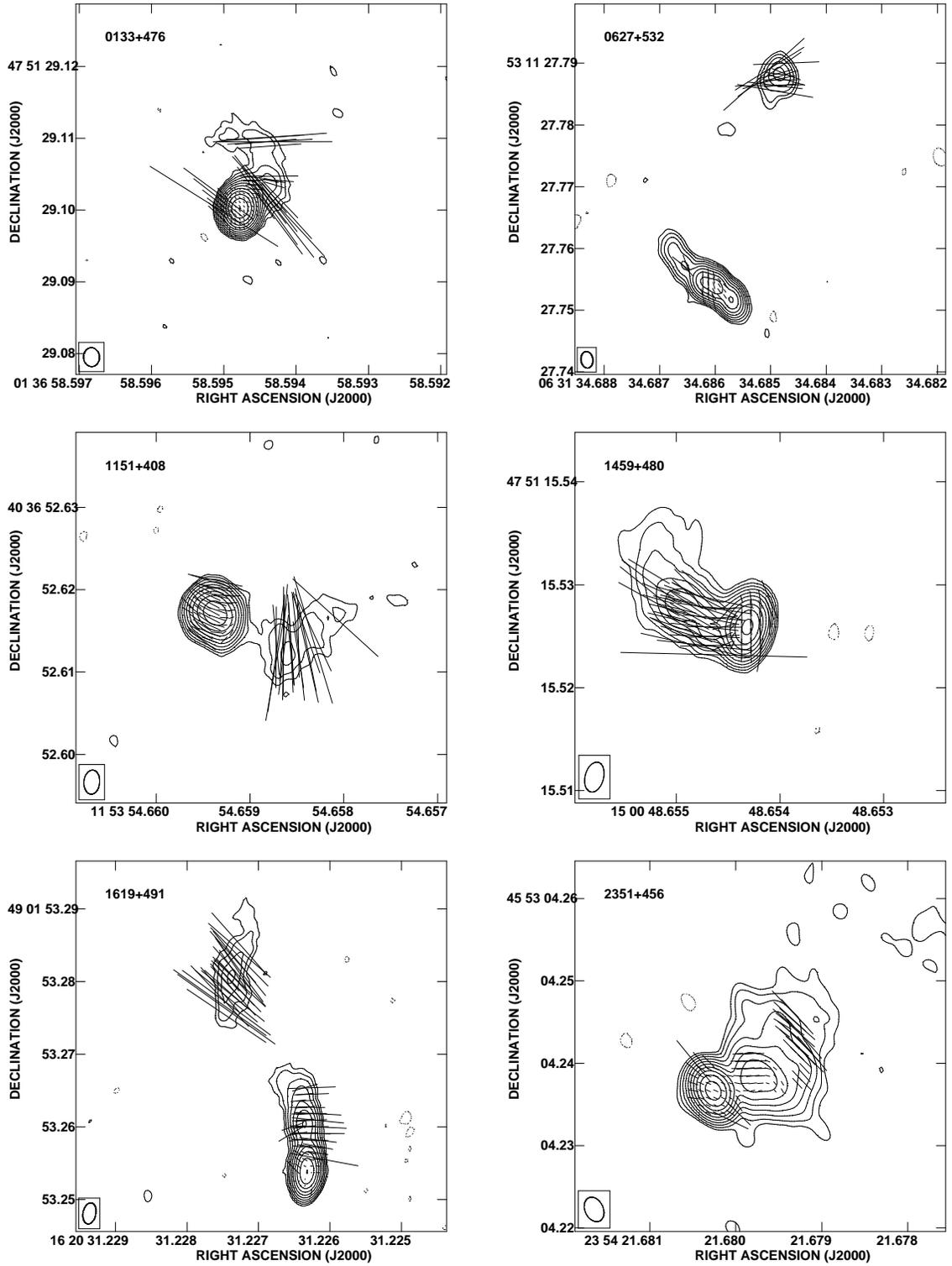}
\caption{A selection of images from the sample showing those sources
with moderate to large bends and some detected polarization.  
Contours, polarization vectors, and blanking is the same as Fig.~\ref{rangemos}.}
\label{bentmos}
\end{figure}

A new, fairly strong correlation between core EVPA and jet axis angle
has been discovered in quasars with projected jet lengths shorter than
6 mas.  These objects exhibit $\bf{E}$-fields nearly perpendicular to
the axis while those quasars with jet lengths longer than 6 mas show
no obvious $\theta-\chi_{\rm core}$ correlation.  That this
correlation exists in spite of the tendancy of quasar cores to have
substantial Faraday RMs (Taylor 1998, 2000), implies that the actual
correlation is even stronger than we observe.  These findings suggest
that quasars with axes anti-aligned with the $\bf{E}$-field are not
intrinsically different from those without this correlation; rather,
the long-jet sources may be at larger angles to the line-of-sight so
that the angle to the core traverses a higher-density region with
magnetic fields that produce sufficient Faraday rotation to smear out
any intrinsic correlation.  This reasoning also can explain the
sparsely populated region of high projected jet lengths and high
fractional core polarizations in Fig.~\ref{lengthstuff}.  These
results can be made consistent with the finding of \citet{list01}, of
electric vectors aligned with the jet axis at 43 GHz if the cores are
optically thick at 5 GHz, but optically thin at 43 GHz.  A magnetic
field perpendicular to the jet direction could be produced by a strong
transverse shock \citep[e.g.][]{laing80}.

Unlike the preferred alignment found in quasar cores and contrary to 
previous claims in the literature
\citep{caw93,hom02}, quasar jets are not found to exhibit a $\theta-\chi$
relationship.  This seems to be the case regardless of jet length, 
core dominance, or fractional polarization. 


In contrast to quasar cores, BL Lac cores in our sample show no
strongly preferred alignment.  Since BL Lacs are observed to have lower
Faraday Rotation Measures than quasars \citep{zav03}, it is unlikely
that the absence of correlation is due to Faraday rotation.  BL Lacs
may be intrinsically different from quasars (i.e., have a lower
power), or their intrinsic high variability may make it impossible to
find a correlation over just one epoch, with only a modest number of
sources (17).  Future multi-epoch analysis, and/or the analysis of a
larger sample, is necessary to conclusively determine the existence of
any preferred alignment.  An interesting test to make in the future
would be to use spectral information to look for preferential
alignments in optically thick or thin cores.  Contrary to claims made
by \citet{gab00} of a bimodal $\theta-\chi_{\rm jet}$ distribution,
the EVPA in BL Lac jets similarly show no correlation with the jet
axis angle.


We found no evolution of source properties (size, fractional polarization,
or polarization angle) with redshift.  This lack
of evolution is 
reasonable if the physical parameters of the central engine outweigh
any effects due to possible variations of intergalactic medium with
redshift.

\acknowledgments

We thank Rene Vermeulen and Silke Britzen for help in scheduling and
for sharing the data.  LKP thanks NSF for summer support through the
Research Experience for Underdergraduates (REU) program.  We thank
Travis Rector and Denise Gabuzda for helpful discussions.  We also
thank an anonymous referee for numerous helpful suggestions.  This
research has made use of the NASA/IPAC Extragalactic Database (NED)
which is operated by the Jet Propulsion Laboratory, California
Institute of Technology, under contract with the National Aeronautics
and Space Administration.

\clearpage

\clearpage

\clearpage

\clearpage

\clearpage

\clearpage

\clearpage

\clearpage

\clearpage

\clearpage

\clearpage

\clearpage

\clearpage

\clearpage

\clearpage

\begin{table}
\begin{tabular}{llllrrrrrrrrrr}
\multicolumn{14}{c}{TABLE 1}  \\
\multicolumn{14}{c}{Measured Source Properties}  \\ \tableline \tableline
\multicolumn{1}{c}{Source} & \multicolumn{1}{c}{E} & \multicolumn{1}{c}{P} & \multicolumn{1}{c}{C} & \multicolumn{1}{c}{$\theta_{axis}$} & \multicolumn{1}{c}{I$_{core}$\tablenotemark{\ast}} & \multicolumn{1}{c}{{\it p}$_{core}$\tablenotemark{\ast}} & \multicolumn{1}{c}{$\chi_{core}$\tablenotemark{\ast}} & \multicolumn{1}{c}{L} & \multicolumn{1}{c}{$\bar{{\it p}}_{jet}$\tablenotemark{\ast}} & \multicolumn{1}{c}{N\tablenotemark{\ast}} & \multicolumn{1}{c}{F\tablenotemark{\ast}} & \multicolumn{1}{c}{$\sigma_{I}$} & \multicolumn{1}{c}{$\sigma_{{\it p}}$}  \\
\multicolumn{1}{c}{(1)} & \multicolumn{1}{c}{(2)} & \multicolumn{1}{c}{(3)} & \multicolumn{1}{c}{(4)} & \multicolumn{1}{c}{(5)} & \multicolumn{1}{c}{(6)} & \multicolumn{1}{c}{(7)} & \multicolumn{1}{c}{(8)} & \multicolumn{1}{c}{(9)} & \multicolumn{1}{c}{(10)} & \multicolumn{1}{c}{(11)} & \multicolumn{1}{c}{(12)} & \multicolumn{1}{c}{(13)} & \multicolumn{1}{c}{(14)} \\ \tableline
0003+380 &            A &            2 & G$_{\rm j}$ &          122 &          545 &    $\leq$1.1 &           -- &         11.6 &           -- &           -- &           -- &         0.18 &         0.10 \\
0016+731 &            C &            P & Q$_{\rm j}$ &            3 &          298 &          3.8 &           79 &          5.1 &           -- &           -- &           -- &         0.22 &         0.11 \\
0035+367 &            C &            E & Q$_{\rm j}$ &           41 &           92 &          0.9 &           25 &          6.0 &         17.3 &           11 &           -- &         0.15 &         0.11 \\
0035+413 &            A &            2 & Q$_{\rm j}$ &          108 &          322 &          2.5 &            2 &         12.9 &         10.3 &           -- &           55 &         0.12 &         0.10 \\
0108+388 &            C &            P & G$_{\rm c}$ &           54 &          146 &    $\leq$1.1 &           -- &          2.7 &           -- &           -- &           -- &         0.28 &         0.11 \\
0109+351 &            A &            2 & Q$_{\rm n}$ &           -- &          369 &          3.7 &           89 &           -- &           -- &           -- &           -- &         0.12 &         0.23 \\
0110+495 &            A &            2 & Q$_{\rm j}$ &          148 &          476 &         16.2 &           31 &         12.6 &         11.0 &           28 &           71 &         0.15 &         0.11 \\
0133+476 &            C &            P & Q$_{\rm j}$ &          129 &         2248 &         23.6 &           33 &          9.8 &           -- &           -- &           -- &         0.26 &         0.14 \\
0145+386 &            A &            2 & Q$_{\rm j}$ &          100 &          286 &          6.5 &           41 &          6.7 &           -- &           -- &           -- &         0.19 &         0.09 \\
0151+474 &            A &            2 & O$_{\rm n}$ &           -- &          451 &         33.2 &          147 &           -- &           -- &           -- &           -- &         0.11 &         0.11 \\
0153+744 &            C &            P & Q$_{\rm j}$ &          113 &          232 &    $\leq$0.7 &           -- &         10.1 &           -- &           -- &           -- &         0.20 &         0.10 \\
0212+735 &            C &            P & BL$_{\rm j}$ &          103 &         2405 &         16.3 &          164 &         13.8 &          9.7 &           59 &           64 &         0.54 &         0.20 \\
0219+428 &            A &            E & BL$_{\rm j}$ &          170 &          612 &         14.2 &          164 &         21.9 &         15.5 &           -- &           33 &         0.15 &         0.13 \\
0227+403 &            A &            2 & Q$_{\rm j}$ &          130 &          285 &         10.8 &          144 &          4.7 &         13.9 &           78 &           -- &         0.11 &         0.10 \\
0249+383 &            A &            2 & Q$_{\rm j}$ &          158 &          345 &          7.5 &           96 &          8.2 &         24.3 &           38 &           21 &         0.12 &         0.09 \\
0251+393 &            A &            2 & Q$_{\rm j}$ &           82 &          217 &          1.8 &          121 &         16.0 &         28.5 &           -- &           39 &         0.10 &         0.10 \\
0256+424 &            B &            2 & O$_{\rm j}$ &           58 &           95 &    $\leq$0.4 &           -- &         10.2 &          8.7 &           66 &           23 &         0.13 &         0.12 \\
0307+380 &            A &            2 & O$_{\rm n}$ &           -- &          707 &         19.6 &          121 &           -- &          8.7 &           -- &           -- &         0.10 &         0.12 \\
0309+411 &            A &            2 & G$_{\rm j}$ &          123 &          238 &    $\leq$0.5 &           -- &         78.1 &           -- &           -- &           -- &         0.11 &         0.09 \\
0316+413 &            C &            P & G$_{\rm j}$ &          178 &         2653 &    $\leq$5.3 &           -- &         11.4 &           -- &           -- &           -- &         3.38 &         0.21 \\
0340+362 &            A &            2 & O$_{\rm n}$ &           -- &          340 &          5.6 &           45 &           -- &           -- &           -- &           -- &         0.10 &         0.10 \\
0402+379 &            B &            1 & G$_{\rm c}$ &          103 &           44 &    $\leq$0.4 &           -- &         21.5 &           -- &           -- &           -- &         0.24 &         0.10 \\
0454+844 &            C &            P & BL$_{\rm n}$ &           -- &          245 &          6.1 &           14 &           -- &           -- &           -- &           -- &         0.11 &         0.10 \\
0537+531 &            A &            2 & Q$_{\rm j}$ &          138 &          507 &          7.0 &          108 &         23.3 &           -- &           -- &           -- &         0.13 &         0.16 \\
0546+726 &            B &            2 & Q$_{\rm j}$ &          121 &          105 &    $\leq$0.3 &           -- &          5.2 &          2.8 &           29 &           -- &         0.18 &         0.10 \\
0554+580 &            A &            2 & Q$_{\rm j}$ &          105 &          197 &          6.1 &          162 &         12.1 &          4.6 &           89 &           -- &         0.11 &         0.10 \\
0600+442 &            A &            2 & G$_{\rm j}$ &          122 &          141 &          0.4 &           86 &         14.1 &          2.7 &           -- &           62 &         0.16 &         0.11 \\
0602+673 &            B &            1 & Q$_{\rm j}$ &           16 &         1010 &          9.3 &           13 &          6.6 &           -- &           -- &           -- &         0.14 &         0.14 \\
0604+728 &            B &            2 & O$_{\rm j}$ &          115 &          131 &    $\leq$0.3 &           -- &         19.9 &           -- &           -- &           -- &         0.27 &         0.10 \\
0609+607 &            A &            2 & Q$_{\rm j}$ &          143 &          572 &         11.3 &           48 &          4.7 &          7.2 &           18 &           -- &         0.13 &         0.12 \\
0620+389 &            B &            1 & Q$_{\rm j}$ &          135 &          350 &         19.7 &            0 &         17.1 &          6.7 &           -- &           73 &         0.12 &         0.11 \\
0621+446 &            A &            E & BL$_{\rm n}$ &           -- &          174 &          4.6 &          136 &           -- &          6.7 &           -- &           -- &         0.09 &         0.10 \\
0627+532 &            A &            2 & Q$_{\rm j}$ &           52 &           81 &    $\leq$0.4 &           -- &         37.4 &          4.2 &           37 &           39 &         0.15 &         0.12 \\
0633+596 &            A &            2 & O$_{\rm j}$ &           41 &          462 &    $\leq$0.9 &           -- &         11.9 &           -- &           -- &           -- &         0.13 &         0.10 \\
0641+393 &            A &            2 & Q$_{\rm j}$ &            0 &          450 &         11.2 &           88 &          5.4 &          6.6 &           83 &           -- &         0.17 &         0.10 \\
0642+449 &            B &            1 & Q$_{\rm j}$ &           86 &         2076 &         18.2 &          175 &          4.5 &           -- &           -- &           -- &         0.18 &         0.28 \\
0646+600 &            A &            1 & Q$_{\rm j}$ &           37 &          640 &    $\leq$1.3 &           -- &          3.0 &           -- &           -- &           -- &         0.19 &         0.11 \\
0650+453 &            A &            2 & Q$_{\rm n}$ &           -- &          193 &          1.8 &          137 &           -- &           -- &           -- &           -- &         0.10 &         0.10 \\
0651+410 &            A &            2 & G$_{\rm n}$ &           -- &          212 &    $\leq$0.4 &           -- &           -- &           -- &           -- &           -- &         0.10 &         0.10 \\
0700+470 &            A &            2 & O$_{\rm j}$ &          112 &          174 &    $\leq$0.3 &           -- &          9.3 &          1.5 &           46 &           -- &         0.11 &         0.09 \\
0702+612 &            A &            2 & Q$_{\rm j}$ &           75 &          167 &          0.3 &          175 &         17.5 &          2.9 &           83 &           -- &         0.12 &         0.10 \\
\tableline
\end{tabular}
\tablenum{1}
\end{table}
\clearpage
\begin{table}
\begin{tabular}{llllrrrrrrrrrr}  \tableline \tableline
\multicolumn{1}{c}{Source} & \multicolumn{1}{c}{E} & \multicolumn{1}{c}{P} & \multicolumn{1}{c}{C} & \multicolumn{1}{c}{$\theta_{axis}$} & \multicolumn{1}{c}{I$_{core}$\tablenotemark{\ast}} & \multicolumn{1}{c}{{\it p}$_{core}$\tablenotemark{\ast}} & \multicolumn{1}{c}{$\chi_{core}$\tablenotemark{\ast}} & \multicolumn{1}{c}{L} & \multicolumn{1}{c}{$\bar{{\it p}}_{jet}$\tablenotemark{\ast}} & \multicolumn{1}{c}{N\tablenotemark{\ast}} & \multicolumn{1}{c}{F\tablenotemark{\ast}} & \multicolumn{1}{c}{$\sigma_{I}$} & \multicolumn{1}{c}{$\sigma_{{\it p}}$}  \\
\multicolumn{1}{c}{(1)} & \multicolumn{1}{c}{(2)} & \multicolumn{1}{c}{(3)} & \multicolumn{1}{c}{(4)} & \multicolumn{1}{c}{(5)} & \multicolumn{1}{c}{(6)} & \multicolumn{1}{c}{(7)} & \multicolumn{1}{c}{(8)} & \multicolumn{1}{c}{(9)} & \multicolumn{1}{c}{(10)} & \multicolumn{1}{c}{(11)} & \multicolumn{1}{c}{(12)} & \multicolumn{1}{c}{(13)} & \multicolumn{1}{c}{(14)}  \\ \tableline
0707+476 &            B &            1 & Q$_{\rm j}$ &           27 &          531 &         29.9 &           62 &         11.8 &           -- &           -- &           -- &         0.10 &         0.10 \\
0710+439 &            C &            P & G$_{\rm c}$ &          173 &          143 &    $\leq$0.9 &           -- &         13.2 &           -- &           -- &           -- &         0.23 &         0.12 \\
0711+356 &            C &            P & Q$_{\rm j}$ &          158 &          597 &         12.8 &          180 &          5.5 &           -- &           -- &           -- &         0.18 &         0.18 \\
0714+457 &            A &            2 & Q$_{\rm j}$ &          128 &          228 &          8.5 &           23 &          4.9 &           -- &           -- &           -- &         0.13 &         0.10 \\
0716+714 &            C &            1 & BL$_{\rm j}$ &            8 &          687 &          6.0 &           19 &         10.3 &           -- &           -- &           -- &         0.13 &         0.09 \\
0724+571 &            A &            2 & Q$_{\rm j}$ &          152 &          316 &          5.2 &           46 &         14.1 &           -- &           -- &           -- &         0.11 &         0.23 \\
0727+409 &            A &            2 & Q$_{\rm j}$ &          131 &          249 &          5.4 &          140 &         15.5 &           -- &           -- &           -- &         0.13 &         0.10 \\
0731+479 &            A &            2 & Q$_{\rm j}$ &           90 &          317 &    $\leq$0.6 &           -- &          3.0 &           -- &           -- &           -- &         0.11 &         0.09 \\
0738+491 &            A &            2 & O$_{\rm n}$ &           -- &          458 &         13.8 &           68 &           -- &          0.9 &           -- &           -- &         0.14 &         0.11 \\
0743+744 &            A &            2 & Q$_{\rm j}$ &           24 &          223 &          3.9 &           82 &          5.2 &         19.1 &           65 &           -- &         0.11 &         0.10 \\
0749+540 &            A &            2 & BL$_{\rm n}$ &           -- &         1559 &         65.4 &           60 &           -- &         19.1 &           -- &           -- &         0.22 &         0.13 \\
0800+618 &            A &            E & O$_{\rm j}$ &          131 &          738 &         17.3 &           50 &          5.5 &           -- &           -- &           -- &         0.16 &         0.12 \\
0803+452 &            A &            2 & Q$_{\rm n}$ &           -- &          257 &          2.5 &           32 &           -- &           -- &           -- &           -- &         0.11 &         0.09 \\
0804+499 &            C &            P & Q$_{\rm n}$ &           -- &          570 &          8.7 &           99 &           -- &           -- &           -- &           -- &         0.15 &         0.12 \\
0814+425 &            C &            P & BL$_{\rm j}$ &          126 &          858 &          9.1 &           85 &          5.6 &           -- &           -- &           -- &         0.31 &         0.14 \\
0824+355 &            A &            2 & Q$_{\rm j}$ &           49 &          420 &          2.4 &          175 &         13.1 &           -- &           -- &           -- &         0.15 &         0.13 \\
0831+557 &            C &            P & G$_{\rm j}$ &          117 &          791 &    $\leq$1.6 &           -- &          9.5 &           -- &           -- &           -- &         1.16 &         0.12 \\
0833+416 &            A &            2 & Q$_{\rm j}$ &            7 &          166 &         21.9 &           88 &          2.4 &          3.4 &           84 &           -- &         0.11 &         0.10 \\
0836+710 &            C &            P & Q$_{\rm j}$ &           36 &         1292 &          3.0 &          111 &         35.2 &         19.1 &           67 &           70 &         0.41 &         0.16 \\
0843+575 &            A &            2 & O$_{\rm j}$ &           37 &           31 &    $\leq$0.3 &           -- &          5.9 &           -- &           -- &           -- &         0.12 &         0.11 \\
0847+379 &            A &            E & G$_{\rm j}$ &            3 &          169 &    $\leq$0.3 &           -- &          6.6 &         21.7 &           -- &           37 &         0.09 &         0.09 \\
0850+581 &            C &            P & Q$_{\rm j}$ &          152 &          600 &          6.3 &            8 &         13.7 &           -- &           -- &           -- &         0.14 &         0.12 \\
0859+470 &            C &            P & Q$_{\rm j}$ &            0 &          371 &          4.2 &          128 &         53.5 &          1.1 &           25 &           -- &         0.20 &         0.14 \\
0900+520 &            A &            2 & Q$_{\rm n}$ &           -- &          165 &          5.8 &          168 &           -- &          1.8 &           -- &           -- &         0.10 &         0.09 \\
0902+490 &            A &            2 & Q$_{\rm n}$ &           -- &          483 &    $\leq$1.0 &           -- &           -- &          1.8 &           -- &           -- &         0.15 &         0.09 \\
0917+624 &            C &            1 & Q$_{\rm j}$ &          162 &         1021 &         39.5 &           40 &          6.0 &           -- &           -- &           -- &         0.78 &         0.13 \\
0923+392 &            C &            P & Q$_{\rm n}$ &           -- &         7902 &         105. &           59 &           -- &           -- &           -- &           -- &         1.49 &         0.34 \\
0925+504 &            A &            2 & BL$_{\rm j}$ &          135 &          415 &         44.2 &           51 &          8.7 &         22.5 &           -- &           15 &         0.14 &         0.10 \\
0927+352 &            A &            2 & Q$_{\rm j}$ &          103 &          242 &          8.7 &           90 &          9.6 &          3.0 &           12 &           -- &         0.12 &         0.10 \\
0929+533 &            A &            2 & Q$_{\rm j}$ &          131 &          241 &          5.2 &           47 &          6.0 &         36.5 &           47 &           -- &         0.11 &         0.09 \\
0930+493 &            A &            2 & Q$_{\rm j}$ &           45 &          260 &         19.0 &          147 &          3.0 &          1.5 &           75 &           -- &         0.12 &         0.10 \\
0942+468 &            A &            2 & G$_{\rm n}$ &           -- &          215 &    $\leq$0.4 &           -- &           -- &          1.5 &           -- &           -- &         0.09 &         0.08 \\
0945+408 &            C &            P & Q$_{\rm j}$ &          127 &         1180 &         28.3 &           31 &         25.0 &         10.1 &           -- &           76 &         0.26 &         0.27 \\
0949+354 &            A &            2 & Q$_{\rm j}$ &          167 &          343 &          4.8 &           47 &          7.7 &          6.7 &           -- &           44 &         0.25 &         0.09 \\
0954+658 &            C &            P & BL$_{\rm j}$ &          106 &          311 &         12.2 &          122 &          9.5 &           -- &           -- &           -- &         0.13 &         0.11 \\
1010+350 &            A &            2 & G$_{\rm j}$ &           95 &          445 &         10.3 &          180 &          8.7 &          9.6 &           54 &           79 &         0.10 &         0.09 \\
1030+398 &            A &            2 & G$_{\rm j}$ &           45 &          372 &    $\leq$0.7 &           -- &          3.0 &           -- &           -- &           -- &         0.10 &         0.10 \\
1031+567 &            C &            P & G$_{\rm c}$ &           49 &            0 &    $\leq$0.8 &           -- &         16.4 &           -- &           -- &           -- &         0.23 &         0.11 \\
1038+528 &            A &            2 & Q$_{\rm j}$ &           27 &          421 &          3.2 &           93 &          4.7 &           -- &           -- &           -- &         0.22 &         0.11 \\
1041+536 &            A &            2 & Q$_{\rm j}$ &          176 &          245 &          3.5 &           64 &          4.8 &         13.7 &           65 &           -- &         0.11 &         0.09 \\
1058+726 &            B &            1 & Q$_{\rm j}$ &            9 &          441 &          2.1 &           61 &         26.8 &          4.6 &           55 &           -- &         0.13 &         0.11 \\
1101+384 &            C &            1 & BL$_{\rm j}$ &          142 &          308 &          6.1 &           35 &         20.4 &           -- &           -- &           -- &         0.15 &         0.14 \\
1105+437 &            A &            2 & Q$_{\rm j}$ &           54 &          261 &          2.9 &           32 &          9.0 &           -- &           -- &           -- &         0.11 &         0.10 \\
\tableline
\end{tabular}
\tablenum{1}
\end{table}
\clearpage
\begin{table}
\begin{tabular}{llllrrrrrrrrrr}  \tableline \tableline
\multicolumn{1}{c}{Source} & \multicolumn{1}{c}{E} & \multicolumn{1}{c}{P} & \multicolumn{1}{c}{C} & \multicolumn{1}{c}{$\theta_{axis}$} & \multicolumn{1}{c}{I$_{core}$\tablenotemark{\ast}} & \multicolumn{1}{c}{{\it p}$_{core}$\tablenotemark{\ast}} & \multicolumn{1}{c}{$\chi_{core}$\tablenotemark{\ast}} & \multicolumn{1}{c}{L} & \multicolumn{1}{c}{$\bar{{\it p}}_{jet}$\tablenotemark{\ast}} & \multicolumn{1}{c}{N\tablenotemark{\ast}} & \multicolumn{1}{c}{F\tablenotemark{\ast}} & \multicolumn{1}{c}{$\sigma_{I}$} & \multicolumn{1}{c}{$\sigma_{{\it p}}$}  \\
\multicolumn{1}{c}{(1)} & \multicolumn{1}{c}{(2)} & \multicolumn{1}{c}{(3)} & \multicolumn{1}{c}{(4)} & \multicolumn{1}{c}{(5)} & \multicolumn{1}{c}{(6)} & \multicolumn{1}{c}{(7)} & \multicolumn{1}{c}{(8)} & \multicolumn{1}{c}{(9)} & \multicolumn{1}{c}{(10)} & \multicolumn{1}{c}{(11)} & \multicolumn{1}{c}{(12)} & \multicolumn{1}{c}{(13)} & \multicolumn{1}{c}{(14)}  \\ \tableline
1106+380 &            A &            E & O$_{\rm j}$ &           31 &           74 &    $\leq$0.5 &           -- &          5.9 &           -- &           -- &           -- &         0.18 &         0.09 \\
1124+455 &            A &            2 & Q$_{\rm j}$ &          175 &          165 &    $\leq$0.4 &           -- &          3.3 &          4.5 &           12 &           -- &         0.12 &         0.13 \\
1144+352 &            A &            2 & G$_{\rm j}$ &          122 &           58 &    $\leq$0.4 &           -- &         23.8 &           -- &           -- &           -- &         0.13 &         0.14 \\
1151+408 &            A &            2 & Q$_{\rm j}$ &           59 &          229 &         19.4 &           68 &         10.5 &         10.8 &           36 &           54 &         0.12 &         0.09 \\
1155+486 &            A &            2 & Q$_{\rm j}$ &           69 &          327 &          9.0 &            0 &          6.7 &           -- &           -- &           -- &         0.15 &         0.11 \\
1205+544 &            A &            2 & O$_{\rm j}$ &          117 &          116 &          0.4 &          105 &          6.7 &           -- &           -- &           -- &         0.12 &         0.10 \\
1206+415 &            A &            2 & BL$_{\rm j}$ &           12 &          162 &          3.0 &           26 &          9.8 &           -- &           -- &           -- &         0.09 &         0.09 \\
1221+809 &            A &            2 & BL$_{\rm j}$ &          173 &          353 &         15.5 &          121 &          9.2 &         16.9 &           37 &           43 &         0.13 &         0.10 \\
1223+395 &            A &            2 & G$_{\rm j}$ &           34 &          118 &    $\leq$0.4 &           -- &         15.4 &          8.1 &           -- &           85 &         0.12 &         0.09 \\
1226+373 &            A &            2 & Q$_{\rm j}$ &          105 &          362 &          1.8 &           25 &          4.7 &           -- &           -- &           -- &         0.09 &         0.09 \\
1239+376 &            A &            2 & Q$_{\rm n}$ &           -- &          253 &         10.2 &          139 &           -- &           -- &           -- &           -- &         0.11 &         0.09 \\
1240+381 &            A &            2 & Q$_{\rm j}$ &           87 &          510 &          3.1 &          160 &         12.4 &           -- &           -- &           -- &         0.20 &         0.09 \\
1250+532 &            A &            2 & BL$_{\rm j}$ &           79 &          174 &          2.1 &           78 &         10.4 &         18.4 &            5 &            1 &         0.13 &         0.09 \\
1258+507 &            A &            2 & Q$_{\rm j}$ &          170 &          184 &          2.4 &           19 &         36.5 &           -- &           -- &           -- &         0.11 &         0.09 \\
1305+804 &            A &            E & O$_{\rm j}$ &           61 &           76 &    $\leq$0.3 &           -- &         13.7 &           -- &           -- &           -- &         0.10 &         0.09 \\
1306+360 &            A &            E & Q$_{\rm n}$ &           -- &          376 &          7.1 &           20 &           -- &           -- &           -- &           -- &         0.10 &         0.14 \\
1308+471 &            A &            2 & Q$_{\rm n}$ &           -- &          220 &          8.6 &          103 &           -- &           -- &           -- &           -- &         0.10 &         0.11 \\
1312+533 &            A &            2 & O$_{\rm j}$ &           63 &          275 &    $\leq$0.6 &           -- &          2.0 &           -- &           -- &           -- &         0.15 &         0.10 \\
1321+410 &            A &            2 & G$_{\rm j}$ &           96 &          158 &    $\leq$0.4 &           -- &          5.4 &           -- &           -- &           -- &         0.19 &         0.10 \\
1325+436 &            A &            2 & Q$_{\rm j}$ &           56 &          230 &          2.1 &           20 &          6.8 &           -- &           -- &           -- &         0.09 &         0.09 \\
1355+441 &            A &            E & G$_{\rm c}$ &          122 &            2 &    $\leq$0.3 &           -- &          5.7 &           -- &           -- &           -- &         0.17 &         0.09 \\
1356+478 &            A &            E & G$_{\rm j}$ &           70 &          121 &    $\leq$0.4 &           -- &          6.1 &           -- &           -- &           -- &         0.12 &         0.09 \\
1413+373 &            A &            2 & Q$_{\rm j}$ &          115 &          180 &          3.5 &           24 &          4.3 &           -- &           -- &           -- &         0.13 &         0.09 \\
1415+463 &            A &            2 & Q$_{\rm j}$ &           81 &          237 &          7.9 &          150 &         10.9 &          6.5 &           57 &           53 &         0.11 &         0.09 \\
1417+385 &            A &            2 & Q$_{\rm n}$ &           -- &         1301 &         47.7 &          124 &           -- &          6.5 &           -- &           -- &         0.12 &         0.13 \\
1421+482 &            A &            2 & Q$_{\rm j}$ &          100 &          108 &          1.8 &           13 &          3.4 &          1.8 &           32 &           -- &         0.09 &         0.09 \\
1424+366 &            A &            2 & Q$_{\rm n}$ &           -- &          406 &          5.8 &           78 &           -- &          1.8 &           -- &           -- &         0.12 &         0.14 \\
1427+543 &            A &            2 & Q$_{\rm j}$ &          135 &          390 &         12.5 &           53 &         10.8 &         10.6 &           64 &           -- &         0.18 &         0.26 \\
1432+422 &            A &            E & O$_{\rm n}$ &           -- &          196 &          6.7 &          179 &           -- &         10.6 &           -- &           -- &         0.11 &         0.09 \\
1448+762 &            A &            2 & G$_{\rm n}$ &           -- &          248 &    $\leq$0.5 &           -- &           -- &         10.6 &           -- &           -- &         0.11 &         0.09 \\
1456+375 &            A &            2 & G$_{\rm j}$ &          108 &          113 &          4.6 &          158 &          2.8 &          8.3 &            1 &           -- &         0.09 &         0.09 \\
1459+480 &            A &            2 & BL$_{\rm j}$ &           86 &          274 &         13.2 &            4 &         10.2 &         17.2 &            0 &           13 &         0.10 &         0.10 \\
1505+428 &            A &            2 & G$_{\rm j}$ &           81 &          420 &    $\leq$0.8 &           -- &          7.6 &           -- &           -- &           -- &         0.11 &         0.10 \\
1534+501 &            A &            2 & Q$_{\rm n}$ &           -- &          133 &          2.5 &          165 &           -- &           -- &           -- &           -- &         0.11 &         0.12 \\
1543+480 &            A &            2 & G$_{\rm j}$ &          114 &          257 &    $\leq$0.5 &           -- &         62.5 &           -- &           -- &           -- &         0.12 &         0.09 \\
1543+517 &            A &            2 & Q$_{\rm j}$ &            0 &          326 &          3.4 &           80 &         24.7 &         12.7 &           68 &           75 &         0.11 &         0.09 \\
1550+582 &            A &            2 & Q$_{\rm n}$ &           -- &          201 &    $\leq$0.4 &           -- &           -- &         12.7 &           -- &           -- &         0.13 &         0.09 \\
1619+491 &            A &            2 & Q$_{\rm j}$ &            3 &          229 &    $\leq$0.5 &           -- &         29.0 &         13.9 &           -- &           67 &         0.12 &         0.10 \\
1622+665 &            A &            E & G$_{\rm n}$ &           -- &          232 &    $\leq$0.5 &           -- &           -- &         13.9 &           -- &           -- &         0.10 &         0.09 \\
1623+578 &            A &            E & O$_{\rm j}$ &           67 &          316 &         18.6 &          110 &         20.9 &          7.2 &           54 &           -- &         0.12 &         0.10 \\
1624+416 &            C &            P & Q$_{\rm j}$ &          172 &          329 &    $\leq$0.7 &           -- &          6.7 &           -- &           -- &           -- &         0.20 &         0.12 \\
1629+495 &            A &            2 & Q$_{\rm j}$ &           66 &          179 &          0.7 &          177 &          5.8 &           -- &           -- &           -- &         0.10 &         0.09 \\
1633+382 &            C &            P & Q$_{\rm j}$ &           98 &         1087 &         12.5 &          164 &          2.1 &          0.8 &           22 &           -- &         0.40 &         0.20 \\
\tableline
\end{tabular}
\tablenum{1}
\end{table}
\clearpage
\begin{table}
\begin{tabular}{llllrrrrrrrrrr}  \tableline \tableline
\multicolumn{1}{c}{Source} & \multicolumn{1}{c}{E} & \multicolumn{1}{c}{P} & \multicolumn{1}{c}{C} & \multicolumn{1}{c}{$\theta_{axis}$} & \multicolumn{1}{c}{I$_{core}$\tablenotemark{\ast}} & \multicolumn{1}{c}{{\it p}$_{core}$\tablenotemark{\ast}} & \multicolumn{1}{c}{$\chi_{core}$\tablenotemark{\ast}} & \multicolumn{1}{c}{L} & \multicolumn{1}{c}{$\bar{{\it p}}_{jet}$\tablenotemark{\ast}} & \multicolumn{1}{c}{N\tablenotemark{\ast}} & \multicolumn{1}{c}{F\tablenotemark{\ast}} & \multicolumn{1}{c}{$\sigma_{I}$} & \multicolumn{1}{c}{$\sigma_{{\it p}}$}  \\
\multicolumn{1}{c}{(1)} & \multicolumn{1}{c}{(2)} & \multicolumn{1}{c}{(3)} & \multicolumn{1}{c}{(4)} & \multicolumn{1}{c}{(5)} & \multicolumn{1}{c}{(6)} & \multicolumn{1}{c}{(7)} & \multicolumn{1}{c}{(8)} & \multicolumn{1}{c}{(9)} & \multicolumn{1}{c}{(10)} & \multicolumn{1}{c}{(11)} & \multicolumn{1}{c}{(12)} & \multicolumn{1}{c}{(13)} & \multicolumn{1}{c}{(14)}  \\ \tableline
1636+473 &            A &            2 & Q$_{\rm j}$ &          156 &          617 &          4.7 &           15 &          8.2 &           -- &           -- &           -- &         0.20 &         0.10 \\
1637+574 &            C &            P & Q$_{\rm j}$ &           23 &          734 &          4.1 &          119 &          3.9 &          5.2 &           64 &           -- &         0.19 &         0.13 \\
1638+540 &            A &            2 & Q$_{\rm n}$ &           -- &          205 &          5.2 &          158 &           -- &          6.6 &           -- &           -- &         0.10 &         0.09 \\
1641+399 &            C &            P & Q$_{\rm j}$ &          114 &         6798 &         75.1 &           73 &         20.2 &           -- &           -- &           -- &         0.97 &         0.77 \\
1642+690 &            C &            P & Q$_{\rm j}$ &           21 &          447 &         13.0 &          133 &         10.3 &          8.8 &           66 &           51 &         0.17 &         0.14 \\
1645+410 &            A &            2 & Q$_{\rm n}$ &           -- &          341 &          1.3 &           58 &           -- &          9.4 &           -- &           -- &         0.11 &         0.10 \\
1652+398 &            A &            P & BL$_{\rm j}$ &          126 &          537 &          1.4 &          166 &         15.1 &         10.2 &           60 &           48 &         0.24 &         0.10 \\
1722+401 &            A &            2 & Q$_{\rm j}$ &          118 &          327 &    $\leq$0.7 &           -- &         12.1 &           -- &           -- &           -- &         0.11 &         0.09 \\
1726+455 &            A &            2 & Q$_{\rm j}$ &           90 &         1075 &         29.5 &          159 &          9.3 &           -- &           -- &           -- &         0.13 &         0.12 \\
1738+499 &            A &            2 & Q$_{\rm j}$ &            0 &          300 &          4.1 &          165 &          6.3 &           -- &           -- &           -- &         0.10 &         0.09 \\
1739+522 &            C &            P & Q$_{\rm j}$ &           52 &          551 &    $\leq$1.1 &           -- &          4.9 &           -- &           -- &           -- &         0.15 &         0.12 \\
1744+557 &            A &            E & G$_{\rm j}$ &           68 &          182 &    $\leq$0.4 &           -- &         15.7 &           -- &           -- &           -- &         0.10 &         0.10 \\
1746+470 &            A &            2 & Q$_{\rm n}$ &           -- &          457 &          2.7 &          180 &           -- &           -- &           -- &           -- &         0.14 &         0.22 \\
1747+433 &            A &            2 & BL$_{\rm j}$ &          168 &          135 &          5.6 &           35 &         13.2 &         28.6 &           -- &           13 &         0.11 &         0.11 \\
1749+701 &            C &            P & BL$_{\rm j}$ &          133 &          359 &          3.0 &           79 &          7.2 &          4.7 &           34 &           -- &         0.15 &         0.12 \\
1755+578 &            A &            2 & Q$_{\rm j}$ &           75 &           58 &    $\leq$0.4 &           -- &         13.2 &           -- &           -- &           -- &         0.11 &         0.10 \\
1803+784 &            C &            P & BL$_{\rm j}$ &           83 &         1780 &         122. &           97 &         26.3 &         22.5 &            5 &           41 &         0.74 &         0.16 \\
1807+698 &            C &            P & G$_{\rm j}$ &           78 &          747 &    $\leq$1.5 &           -- &         41.6 &         11.9 &           -- &           15 &         0.18 &         0.13 \\
1812+412 &            A &            2 & Q$_{\rm j}$ &           82 &          222 &          3.8 &          132 &         14.3 &         10.2 &           -- &           84 &         0.11 &         0.10 \\
1823+568 &            C &            P & BL$_{\rm j}$ &           18 &          629 &         16.7 &           12 &         25.3 &         20.3 &           -- &           12 &         0.16 &         0.15 \\
1828+399 &            A &            2 & O$_{\rm n}$ &           -- &          157 &    $\leq$0.3 &           -- &           -- &         13.0 &           -- &           -- &         0.09 &         0.09 \\
1839+389 &            A &            2 & Q$_{\rm n}$ &           -- &          196 &          1.6 &           95 &           -- &         13.0 &           -- &           -- &         0.09 &         0.09 \\
1842+681 &            A &            1 & Q$_{\rm j}$ &          131 &          342 &          5.2 &           21 &         11.5 &          5.4 &           86 &           -- &         0.14 &         0.14 \\
1850+402 &            A &            2 & Q$_{\rm j}$ &           63 &          407 &          7.1 &           72 &          2.0 &           -- &           -- &           -- &         0.13 &         0.10 \\
1851+488 &            A &            2 & Q$_{\rm n}$ &           -- &          248 &          5.0 &          144 &           -- &           -- &           -- &           -- &         0.12 &         0.13 \\
1908+484 &            A &            2 & O$_{\rm j}$ &           60 &           96 &    $\leq$0.3 &           -- &         11.8 &           -- &           -- &           -- &         0.10 &         0.10 \\
1910+375 &            A &            2 & Q$_{\rm j}$ &          176 &          268 &         11.7 &           73 &          8.1 &         20.1 &           -- &           76 &         0.13 &         0.83 \\
1924+507 &            A &            2 & Q$_{\rm j}$ &            0 &          217 &          1.7 &            6 &          9.3 &          2.2 &           36 &           -- &         0.11 &         0.11 \\
1928+738 &            C &            P & Q$_{\rm j}$ &          159 &         1664 &         15.9 &           79 &         20.2 &          8.6 &           80 &           53 &         0.66 &         0.18 \\
1943+546 &            A &            1 & G$_{\rm c}$ &           84 &           14 &    $\leq$0.6 &           -- &         25.5 &           -- &           -- &           -- &         0.24 &         0.09 \\
1946+708 &            A &            2 & G$_{\rm c}$ &           32 &           29 &    $\leq$0.3 &           -- &         18.1 &           -- &           -- &           -- &         0.31 &         0.09 \\
1954+513 &            C &            P & Q$_{\rm j}$ &          125 &          580 &          1.4 &           45 &         14.8 &           -- &           -- &           -- &         0.18 &         0.12 \\
2021+614 &            C &            P & Q$_{\rm c}$ &           41 &           91 &    $\leq$2.5 &           -- &          6.4 &           -- &           -- &           -- &         0.36 &         0.18 \\
2054+611 &            A &            2 & O$_{\rm j}$ &          165 &          253 &          3.0 &           12 &          7.1 &          7.0 &           62 &           83 &         0.12 &         0.11 \\
2116+818 &            A &            E & G$_{\rm j}$ &          153 &           72 &    $\leq$0.3 &           -- &          8.2 &           -- &           -- &           -- &         0.12 &         0.09 \\
2136+824 &            A &            2 & Q$_{\rm j}$ &          139 &          117 &          2.7 &          128 &         35.3 &          5.1 &           40 &           40 &         0.18 &         0.12 \\
2138+389 &            B &            2 & Q$_{\rm j}$ &           90 &          124 &    $\leq$0.3 &           -- &          6.0 &           -- &           -- &           -- &         0.13 &         0.11 \\
2200+420 &            C &            P & BL$_{\rm j}$ &          162 &         1098 &         19.5 &          170 &         15.2 &           -- &           -- &           -- &         0.35 &         0.16 \\
2214+350 &            A &            1 & Q$_{\rm j}$ &            7 &          346 &          1.9 &          173 &          7.3 &           -- &           -- &           -- &         0.12 &         0.10 \\
2235+731 &            A &            2 & O$_{\rm j}$ &           32 &          287 &          1.1 &          106 &          3.9 &          8.0 &           57 &           -- &         0.11 &         0.10 \\
2238+410 &            A &            2 & Q$_{\rm j}$ &          133 &          320 &          7.2 &           11 &          4.9 &           -- &           -- &           -- &         0.11 &         0.10 \\
2259+371 &            A &            2 & Q$_{\rm j}$ &           11 &          225 &          7.5 &          108 &          4.9 &         10.1 &           14 &           -- &         0.12 &         0.10 \\
2309+454 &            A &            2 & O$_{\rm j}$ &          114 &          313 &          0.7 &           21 &          3.0 &          6.4 &           48 &           -- &         0.10 &         0.10 \\
\tableline
\end{tabular}
\tablenum{1}
\end{table}
\clearpage
\begin{table}
\begin{tabular}{llllrrrrrrrrrr}  \tableline \tableline
\multicolumn{1}{c}{Source} & \multicolumn{1}{c}{E} & \multicolumn{1}{c}{P} & \multicolumn{1}{c}{C} & \multicolumn{1}{c}{$\theta_{axis}$} & \multicolumn{1}{c}{I$_{core}$\tablenotemark{\ast}} & \multicolumn{1}{c}{{\it p}$_{core}$\tablenotemark{\ast}} & \multicolumn{1}{c}{$\chi_{core}$\tablenotemark{\ast}} & \multicolumn{1}{c}{L} & \multicolumn{1}{c}{$\bar{{\it p}}_{jet}$\tablenotemark{\ast}} & \multicolumn{1}{c}{N\tablenotemark{\ast}} & \multicolumn{1}{c}{F\tablenotemark{\ast}} & \multicolumn{1}{c}{$\sigma_{I}$} & \multicolumn{1}{c}{$\sigma_{{\it p}}$}  \\
\multicolumn{1}{c}{(1)} & \multicolumn{1}{c}{(2)} & \multicolumn{1}{c}{(3)} & \multicolumn{1}{c}{(4)} & \multicolumn{1}{c}{(5)} & \multicolumn{1}{c}{(6)} & \multicolumn{1}{c}{(7)} & \multicolumn{1}{c}{(8)} & \multicolumn{1}{c}{(9)} & \multicolumn{1}{c}{(10)} & \multicolumn{1}{c}{(11)} & \multicolumn{1}{c}{(12)} & \multicolumn{1}{c}{(13)} & \multicolumn{1}{c}{(14)}  \\ \tableline
2310+385 &            A &            2 & Q$_{\rm j}$ &           60 &          202 &          2.1 &          102 &          4.8 &          2.7 &           53 &           -- &         0.10 &         0.11 \\
2319+444 &            A &            2 & Q$_{\rm n}$ &           -- &          303 &         15.7 &          134 &           -- &          2.7 &           -- &           -- &         0.11 &         0.10 \\
2346+385 &            A &            2 & Q$_{\rm j}$ &          149 &          344 &         10.2 &           51 &          5.2 &           -- &           -- &           -- &         0.14 &         0.10 \\
2351+456 &            C &            P & Q$_{\rm j}$ &          108 &          768 &         14.8 &           66 &          9.8 &          4.5 &           14 &           -- &         0.25 &         0.18 \\
2352+495 &            A &            P & G$_{\rm c}$ &            2 &            8 &    $\leq$0.9 &           -- &         25.5 &           -- &           -- &           -- &         0.33 &         0.09 \\
2356+385 &            A &            2 & Q$_{\rm n}$ &           -- &          329 &          3.6 &           24 &           -- &           -- &           -- &           -- &         0.12 &         0.10 \\
2356+390 &            A &            2 & Q$_{\rm j}$ &           51 &          125 &          8.5 &          164 &         10.8 &          7.5 &            0 &           28 &         0.09 &         0.09 \\

\tableline
\end{tabular}
\tablenum{1}
\tablenotetext{\ast}{A more complete table with all measured source
  properties, including jet components, can be found at {\tt www.aoc.nrao.edu/$\sim$gtaylor/cjftab.text}.}
\tablecomments{Col. (1): B1950 source name.  Col. (2): E, epoch -- A:1998; B:1999; C:2000.  Col. (3): P, parent sample -- P:Pearson \& Readhead 1988; 1: first Caltech-Jodrell Bank survey (CJ1, Polatidis et al.\ 1995); 2: second Caltech-Jodrell Bank survey (CJ2, Taylor et al.\ 1994); E: 18 sources not included in PR, CJ1 or CJ2 that complete the Caltech-Jodrell Bank flat-spectrum survey (CJF).  Col. (4): C, class.  Key to identifications: Q:quasar; G:galaxy; BL:BL Lac object; O:other.  Key to subscripts: j:core-jet morphology; n:naked core morphology; c:Compact Symmetric Object.  Col. (5): jet axis angle (deg). Col. (6): total intensity at the core (mJy).  Col. (7): total polarized intensity at the core (mJy); cores with detected polarized intensities have 1$\sigma_{\rm typical}\approx0.2$ mJy.  Col. (8): electric vector position angle at the core (deg); cores that meet detection criteria have $1\sigma_{\rm typical}\approx4\dg$.  Col. (9): L,jet length (mas).  Col. (10): average polarized intensity of detected jet components (mJy).  Col. (11): N, average difference between jet axis angle and electric vector position angle in detected jet components within 6 mas of the core (deg).  Col. (12): F, average difference between jet axis angle and electric vector position angle in detected jet components farther than 6 mas from the core (deg).  Col. (13): total intensity RMS noise (mJy).  Col. (14): polarized intensity RMS noise (mJy).}
\end{table}
\clearpage


\clearpage

\begin{thebibliography}{}
\bibitem[Antonucci(1993)]{ant93} Antonucci, R. 1993, ARA\&A, 31, 473
\bibitem[Britzen et al. (1998)]{Britz98} Britzen, S., Vermeulen, R.~C., 
Taylor, G.~B., 
Readhead, C.~S., Pearson, T.~J., Henstock, D.~R., \& Wilkinson, P.~N.  
22-26 June, 1998, in BL Lac Phenomenon conference held in Turku, Finland, p. 431
\bibitem[Cao \& Jiang (2002)]{cj02} Cao, X.~\& Jiang, D.~R.\ 
2002, \mnras, 331, 111 
\bibitem[Cawthorne et al.(1993)]{caw93} Cawthorne, T.~V., Wardle, J.~F.~C., 
Roberts, D.~H., Gabuzda, D.~C., \& Brown, L.~F.\ 1993, \apj, 416, 519 
\bibitem[Feigelson \& Nelson(1985)]{fn85} Feigelson, 
E.~D.~\& Nelson, P.~I.\ 1985, \apj, 293, 192 
\bibitem[Gabuzda, Pushkarev \& Cawthorne(2000)]{gab00} Gabuzda, D.C., 
Pushkarev, A.B., \& Cawthorne, T.V. 2000, \mnras, 319, 1109
\bibitem[Gabuzda (2002)]{gab02} Gabuzda, D.C.\ 2002, Bologna Jets
Workshop, in press
Pushkarev, A.B., \& Cawthorne, T.V. 2000, \mnras, 319, 1109
\bibitem[G{\' o}mez et al.(2000)]{jlg00} G{\' o}mez, J.~L., 
Marscher, A.~P., Alberdi, A., Jorstad, S.~G., \& Garc\'{\i}a-Mir{\' o}, C.\ 
2000, Science, 289, 2317
\bibitem[Gurvits, Kellermann, \& Frey(1999)]{gur99} Gurvits, 
L.~I., Kellermann, K.~I., \& Frey, S.\ 1999, \aap, 342, 378
\bibitem[Henstock et al.(1995)]{hen95} Henstock, D.~R., 
Browne, I.~W.~A., Wilkinson, P.~N., Taylor, G.~B., Vermeulen, R.~C., 
Pearson, T.~J., \& Readhead, A.~C.~S.\ 1995, \apjs, 100, 1 
\bibitem[Homan et al.(2002)]{hom02} Homan, D.~C., Ojha, R., 
Wardle, J.~F.~C., Roberts, D.~H., Aller, M.~F., Aller, H.~D., \& Hughes, 
P.~A.\ 2002, \apj, 568, 99 
\bibitem[Isobe, Feigelson, \& Nelson(1986)]{ifn86} Isobe, T., 
Feigelson, E.~D., \& Nelson, P.~I.\ 1986, \apj, 306, 490
\bibitem[Kellermann (1993)]{kel93} Kellermann, K.~I.\ 1993, 
\nat, 361, 134 
\bibitem[Lavalley, Isobe, \& Feigelson(1992)]{lif92} 
Lavalley, M., Isobe, T., \& Feigelson, E.\ 1992, Astronomical Data Analysis 
Software and Systems I, A.S.P.~Conference Series, Vol.~25, 1992, Diana 
M.~Worrall, Chris Biemesderfer, and Jeannette Barnes, eds., p.~245., 1, 245 
\bibitem[Laing (1980)]{laing80} Laing, R.~A.\ 1980, MNRAS, 193, 439
\bibitem[Lister \& Marscher(1997)]{lis97} Lister, M.~L.~\& 
Marscher, A.~P.\ 1997, ApJ, 476, 572 
\bibitem[Lister (2001)]{list01} Lister, M.~L.\ 2001, ApJ, 562, 208
\bibitem[March{\~ a} \& Browne (1995)]{mar95} March{\~ a}, M.~J.~M.\ 1995, 
MNRAS, 275, 951
\bibitem[Meier (2001)]{mei01} Meier, D.~L., Koide, S., \& Uchida, Y.\ 
2001, Science, 291, 84
\bibitem[Owsianik \& Conway(1988)]{ows98} Owsianik, I., \& Conway, J.~E.\ 
1998, A\&A, 337, 69
\bibitem[Roberts et al.(1984)]{rob84} Roberts, D.~H., Potash, 
R.~I., Wardle, J.~F.~C., Rogers, A.~E.~E., \& Burke, B.~F.\ 1984, IAU 
Symp.~110: VLBI and Compact Radio Sources, 110, 35 
\bibitem[Roberts, Brown, \& Wardle(1991)]{rob91} Roberts, 
D.~H., Brown, L.~F., \& Wardle, J.~F.~C.\ 1991, ASP Conf.~Ser.~ 19: IAU 
Colloq.~131: Radio Interferometry.~Theory, Techniques, and Applications, 
281 
\bibitem[Pearson \& Readhead(1988)]{pr88} Pearson, T.~J., \& Readhead, 
A.~C.~S.\ 1988, ApJ, 328, 114
\bibitem[Peck \& Taylor (2000)]{peck00} Peck, A.~B., \& Taylor, G.~B.\
2000, ApJ, 533, 95
\bibitem[Polatidis et al.(1995)]{polcj1} Polatidis, A.~G., 
Wilkinson, P.~N., Xu, W., Readhead, A.~C.~S., Pearson, T.~J., Taylor, 
G.~B., \& Vermeulen, R.~C.\ 1995, ApJS, 98, 1 
\bibitem[Press et al.(1992)]{numrec} Press, W.~H., Teukolsky, S.~A., 
Vetterling, W.~T., \& Flannery, B.~P. 1992, Numerical Recipes in C: 
The Art of Scientific Computing (2nd ed.; Cambridge: Cambridge 
University Press)
\bibitem[Readhead et al.(1996)]{rea96} Readhead, A. C. S., Taylor,
G. B., Xu, W., Pearson, T. J., Wilkinson, P. N., \& Polatidis,
A. G. 1996, ApJ, 460, 612
\bibitem[Schwab \& Cotton(1983)]{sc83} Schwab, F.~R., \& Cotton, W.~D.~1983, AJ, 
88, 688
\bibitem[Shepherd, Pearson \& Taylor (1995)]{Shep95} Shepherd, M.~C., Pearson, T.~J., 
\& Taylor, G.~B.~1994, BAAS, 26, 987
\bibitem[Taylor et al.(1994)]{taycj2} Taylor, G.~B., 
Vermeulen, R.~C., Pearson, T.~J., Readhead, A.~C.~S., Henstock, D.~R., 
Browne, I.~W.~A., \& Wilkinson, P.~N.\ 1994, ApJS, 95, 345 
\bibitem[Taylor, Readhead \& Pearson (1996)]{trp96} Taylor, G.B., 
Readhead, A.C.S., \& Pearson, T.J. 1996, \apj, 463, 95
\bibitem[Taylor et al.(1996)]{TayCJF} Taylor, G.~B., Vermuelen, R.C., Readhead, 
A.~C.~S., Pearson, T.~J., Henstock, D.R., \& Wilkinson, P.N. 1996, \apjs, 107, 37
\bibitem[Taylor \& Vermeulen (1997)]{Tay97} Taylor, G.~B., \ Vermeulen, R.~C.\ 1997, 
ApJ, 485, L9
\bibitem[Taylor (1998)]{tay98} Taylor, G.~B.  1998, \apj, 506, 637
\bibitem[Taylor (2000)]{tay00} Taylor, G.~B.  2000, \apj, 533, 95
\bibitem[Taylor \& Myers (2000)]{tmy00} Taylor, G.~B. \& Myers, S.~T. 2000 
VLBA Scientific Memo 26, National Radio Astronomy Observatory
\bibitem[Wilkinson et al.(2001)]{wil01} Wilkinson, P.~N.~et 
al.\ 2001, Physical Review Letters, 86, 584 
\bibitem[Vermeulen et al.(1995)]{ver95} Vermeulen, R.~C., 
Ogle, P.~M., Tran, H.~D., Browne, I.~W.~A., Cohen, M.~H., Readhead, 
A.~C.~S., Taylor, G.~B., \& Goodrich, R.~W.\ 1995, ApJ, 452, L5 
\bibitem[Vermeulen, Taylor, Readhead, \& Browne(1996)]{ver96} 
Vermeulen, R.~C., Taylor, G.~B., Readhead, A.~C.~S., \& Browne, I.~W.~A.\ 
1996, \aj, 111, 1013 
\bibitem[Zavala \& Taylor (2001)]{zt01} Zavala, R.~T. \& Taylor, G.~B. 2001, 
\apj, 550, L147
\bibitem[Zavala \& Taylor (2003)]{zav03} Zavala, R.~T. \& Taylor, G.~B. 2003, 
\apj, in press
\end{thebibliography}
\end{document}